\documentclass{aa}

\usepackage{newtxtext,newtxmath}

\usepackage[T1]{fontenc}
\usepackage{ae,aecompl}

\usepackage{natbib}
\bibpunct{(}{)}{;}{a}{}{,} 

\usepackage{graphicx}	
\usepackage{amsmath}	
\usepackage{amssymb}	
\usepackage{xcolor}
\usepackage[normalem]{ulem}
\usepackage{subcaption}

\providecommand{\gaia}{\textit{Gaia }}
\providecommand{\gaianospace}{\textit{Gaia}}
\providecommand{\kms}{km s$^{-1}$ }
\providecommand{\degr}{$^{\circ}$ }

\begin{document}

\title{Kinematic analysis of the Large Magellanic Cloud \\
using Gaia DR3\thanks{The LMC / MW classification probability of each object will be made available in electronic form at the CDS via anonymous ftp to cdsarc.u-strasbg.fr (130.79.128.5) or via http://cdsweb.u-strasbg.fr/cgi-bin/qcat?J/A+A/}}

\author{Ó. Jiménez-Arranz\inst{1,2,3}
   \and M. Romero-Gómez\inst{1,2,3}
   \and X. Luri\inst{1,2,3}
   \and P. J. McMillan\inst{4}
   \and T. Antoja\inst{1,2,3}
   \and L. Chemin\inst{5}
   \and \\ S. Roca-Fàbrega\inst{6,7}
   \and E. Masana\inst{1,2,3}
   \and A. Muros\inst{1,2}}

\institute{{Departament de Física Quàntica i Astrofísica (FQA), Universitat de Barcelona (UB), C Martí i Franquès, 1, 08028 Barcelona, Spain}
\and
{Institut de Ciències del Cosmos (ICCUB), Universitat de Barcelona, Martí i Franquès 1, 08028 Barcelona, Spain}
\and
{Institut d’Estudis Espacials de Catalunya (IEEC), C Gran Capità, 2-4, 08034 Barcelona, Spain}
\and
{Lund Observatory, Department of Astronomy and Theoretical Physics, Lund University, Box 43, 22100 Lund, Sweden}
\and
{Centro de Astronomía - CITEVA, Universidad de Antofagasta, Avenida Angamos 601, Antofagasta 1270300, Chile}
\and
{Dpto. Física de la Tierra y Astrofísica, Universidad Complutense de Madrid, Madrid, Spain}
\and
{Instituto de Astronomía, Universidad Nacional Autónoma de México, Apartado Postal 106, C. P. 22800, Ensenada, B. C., Mexico }}

\date{Received <date> / Accepted <date>}

\abstract 
{The high quality of the \gaia mission data is allowing to study the internal kinematics of the Large Magellanic Cloud (LMC) in unprecedented detail, providing insights on the non-axisymmetric structure of its disc. Recent works by the \gaia Collaboration have already made use of the excellent proper motions of \gaianospace~DR2 and \gaianospace~EDR3 for a first such analysis, but were based on limited strategies to distinguish the LMC stars from the Milky Way foreground that do not use all the available information and could not use the third component of the stellar motion, the line of sight velocity, that is now available on \gaianospace~DR3 for a significant number of stars.}
{Our aim is twofold: First, to define and validate an improved, more efficient and adjustable selection strategy to distinguish the LMC stars from the Milky Way foreground. Second, to check the possible biases that assumed parameters or sample contamination from the Milky Way can introduce in the analysis of the internal kinematics of the LMC using \gaia data.} 
{Our selection is based on a supervised Neural Network classifier using as much as of the \gaianospace~DR3 data as possible. Using this classifier we select three samples of candidate LMC stars with different degrees of completeness and purity; we validate these classification results using different test samples and we compare them with the results from the selection strategy used in the \gaia Collaboration papers, based only on proper motions. We analyse the resulting velocity profiles and maps for the different LMC samples, and we check how these results change when using also the line-of-sight velocities, available for a subset of stars.} 
{We show that the contamination in the samples from Milky Way stars affects basically the results for the outskirts of the LMC, and that the analysis formalism used in absence of line-of-sight velocities does not bias the results for the kinematics in the inner disc. For the first time, we perform a kinematic analysis of the LMC using samples with the full three dimensional velocity information from \gaianospace~DR3.}
{The detailed 2D and 3D kinematic analysis of the LMC internal dynamics show that: the dynamics in the inner disc is mainly bar dominated; the kinematics on the spiral arm over-density seem to be dominated by an inward motion and a rotation faster than that of the disc in the piece of the arm attached to the bar; contamination of MW stars seem to dominate the outer parts of the disc and mainly affects old evolutionary phases; uncertainties in the assumed disc morphological parameters and line-of-sight velocity of the LMC can in some cases have significant effects in the analysis.}


\maketitle

\section{Introduction}

The Large Magellanic Cloud (LMC) is one of the Milky Way (MW) satellite galaxies and member of the Local Group. 
The LMC is the prototype of dwarf, bulgeless spiral galaxy (the so-called Magellanic type, Sm), with an asymmetric stellar bar, many star forming regions, including the Tarentula Nebula, and prominent spiral arms  \citep[e.g.][]{Elmegreen1980,Gallagher1984,Yozin2014, luri20}. The LMC is a gas-rich galaxy characterised by an inclined disc \citep[e.g.][]{vdm01,vandermarel01}, with several warps \citep[e.g.][]{Olsen2002,Choi2018,ripepi22} and an offset bar whose origin is not well understood \citep[e.g.][]{Zaritsky2004}.
Due to its proximity, the LMC is a perfect target for many studies and focused photometric surveys, such as VMC-VISTA Survey of the Magellanic Clouds system \citep{cioni11} or SMASH-Survey of the Magellanic Stellar History \citep{Nidever2017}, as well as the astrometric mission \gaia (ESA). Already in \citet{helmi18} and \citet{luri20} (hereafter, MC21), the authors show the capabilities of \gaia to characterise the structure and kinematics of this nearby galaxy. The recovery of its three-dimensional structure using \gaia data only has been shown to be complex due to the zero point in parallax and limit in parallax uncertainties \citep[MC21, ][]{Lindegren2021a,Lindegren2021b}. Recent attempts using  specific populations for which individual distances can be anchored such as Cepheids \citep{ripepi22} or RR Lyrae \citep{cusano21} have been more effective. Three-dimensional structure analysis is not the only tool to infer the characteristics and morphology of the galaxy under study. Kinematic profiles and kinematic maps add information on the characteristics and dynamical evolution of the galaxy \citep[] [MC21]{helmi18,Vasiliev2018}. 

The presence of non-axisymmetric features, such as a bar or spiral arms, modifies the velocity map of a simply rotating disc. The nature of the spiral arms, whether it is a density wave \citep{Lyndblad1960,LinShu1964}, a tidally induced arm \citep[e.g.][]{ToomreToomre1972}, transient co-rotating arms \citep{GoldreichLyndenBell1965,JulianToomre1966,Toomre1981} or bar induced \citep[e.g.][]{Athanassoula1980,RomeroGomez2007,Salo2010,Garma2021}, can be disentangled from its signature in the velocity field \citep[e.g.][]{rocafabrega2013,RocaFabrega2014}. How the LMC became so asymmetric, particularly with regard to the ultimate origin of its spiral arm in contrast to more massive spiral galaxies, remains unclear, and detailed kinematic profiles and maps are necessary to investigate.

Kinematics of stars in the outskirts of the LMC have shed some light on the characteristics of the stellar bridge generated by the tidal interaction between the Large and Small Magellanic Clouds \citep{Zivick2019,Schmidt20}, or the formation of the LMC's northern arm \citep{Cullinane2022a} or the dynamical equilibrium of the disc \citep{Cullinane2022b}. Pre-\gaia proper motions and line-of-sight velocities of less than a thousand stars where used by \citet{Kallivayalil2013,vandermarel&kallivayalil2014} to show the detailed large-scale rotation of the LMC disc. The number of sources increased by orders of magnitude when using \gaianospace~DR2 proper motions to study the internal motion of the LMC \citep{helmi18,Vasiliev2018}. \citet{Wan2020} used Carbon Stars and \gaianospace~DR2 proper motions to infer the LMC centre, systemic motion and morphological parameters, and compare them with other stellar populations. Similarly, the improved accuracy of \gaianospace~EDR3 \citep[MC21,][]{Niederhofer2022} allowed a detailed study of the LMC disc kinematics, to separate the analysis by different stellar evolutionary phases, and to extend the study to the LMC outskirts and bridge between the LMC and SMC. 

In this work we focus on the general kinematic analysis of the LMC disc and we present the first 3D velocity maps and profiles of the LMC measured using \gaianospace~DR3 proper motions and line-of-sight velocities. It is the first time that a homogeneous data set of a galaxy that is not the Milky Way with 3D velocity information is presented, for more than 20 thousand stars. We compare the maps with the ones obtained from previous \gaia releases where only astrometric motions were considered. With the new maps, we thus want to assess where, and to which extent, the kinematics have benefited from the line-of-sight velocities.

This paper is organised as follows. In Section~\ref{sec:data} we describe the LMC samples used throughout this work. We use a new supervised classification strategy based on Neural Networks to separate the LMC stars from the MW foreground stars. In this section we also describe the training sample, how we apply the classifier to \gaia data, and how we validate the classification. In Section~\ref{sec:trans} we describe the formalism adopted to transform from \gaia observables to the LMC reference frame, and its validation using an N-body simulation. In Section~\ref{sec:velocity} we show the detailed kinematic analysis of the LMC samples, showing the velocity profiles and the velocity maps of the different LMC samples. In Section~\ref{sec:discussion} we study possible biases on the velocity maps caused by the unknown LMC 3D geometry, and uncertainties in the systemic motion. Finally, in Section~\ref{sec:conclusions} we summarise the main conclusions of this work.

\section{Data selection}
\label{sec:data}

In this section we describe the method to select the samples of stars used in this paper. Our starting point is the base sample obtained by selecting \gaianospace~DR3 \citep{gaiadr3} stars around the center of the LMC. This base sample is a mixture of MW foreground stars and LMC stars. Ideally, one could distinguish both types of objects through their distances, but due to the large uncertainties in the parallax-based distances at LMC \citep[MC21, ][]{edr3_astrometric}, a selection of LMC sources exclusively based on parallaxes is not possible and would be efficient only to remove bright MW stars.

Therefore, in order to build a sample of LMC stars for the kinematic analysis in this paper, we need to define a selection criteria to separate them from the MW foreground. A first option is to use a proper motion based selection (Section~\ref{sec:PM}) as done in MC21; we have kept this methodology to provide a common reference with the results in that paper. We have also implemented an alternative selection method based on machine learning classifiers (Neural Networks, see Section~\ref{sec:NN}) because, firstly, a selection purely based on proper motions might have some effect on the kinematic analysis and, secondly, we wanted to use the full data available in the Gaia catalogue to improve the classification.

We have created the following working samples: 

\begin{itemize}
\item Based on \gaia data:
\begin{itemize}
\item \gaia base sample: initial \gaia DR3 sample selected around the LMC center, before applying any further cut or classification. Described in Section~\ref{sec:gaiabase}
\item \gaia LMC Proper Motion (PM) sample: Application of a proper motion cut to the \gaia base sample. Described in Section~\ref{sec:PM}.
\item \gaia LMC Complete, Optimal and Truncated-Optimal samples: the result of the Neural Network (NN) classification. Described in Section~\ref{sec:NNapplication}.
\item Validation samples, described in Section~\ref{sec:NNvalidation}:
\begin{itemize}
\item LMC Cepheids
\item LMC RR-Lyrae
\item LMC+MW StarHorse
\end{itemize}
\end{itemize}
\item Based on simulations:
\begin{itemize}
\item \gaia (MW+LMC) training sample: simulation based on the \gaia Object Generator (GOG). Described in Section~\ref{sec:NNtrainingsample}.
\end{itemize}
\end{itemize}

\subsection{\gaia base sample}
\label{sec:gaiabase}

The Gaia base sample was obtained using a selection from the \texttt{gaia\_source} table in \gaia DR3 with a $15^{\circ}$ radius around the LMC centre defined as $(\alpha, \delta) = (81.28^{\circ}$, $69.78^{\circ})$ \citep{vandermarel01} and a limiting $G$ magnitude of $20.5$. We only keep stars with parallax and integrated photometry information since they are used in the LMC/MW classification. This selection can be reproduced using the following ADQL query in the \gaia archive:

\begin{verbatim}
SELECT * FROM gaiadr3.gaia_source as g
WHERE 1=CONTAINS(POINT('ICRS',g.ra,g.dec),
CIRCLE('ICRS',81.28,-69.78,15))
AND g.parallax IS NOT NULL
AND g.phot_g_mean_mag IS NOT NULL
AND g.phot_bp_mean_mag IS NOT NULL
AND g.phot_rp_mean_mag IS NOT NULL
AND g.phot_g_mean_mag < 20.5
\end{verbatim}

The resulting base sample contains a total of 18~783~272 objects.

\subsection{Proper motion based classification}
\label{sec:PM}

We use the same selection based on the proper motions of the stars as in MC21 to provide a baseline comparison with these previous results.  In short, the median proper motions of the LMC are determined from a sample restricted to its very centre, minimising the foreground contamination by a cut in magnitude and parallax. We keep only stars whose proper motions obey the constraint $\chi^2 < 9.21$, i.e., an estimated 99\% confidence region (see details in Section~2.2 of MC21). The resulting sample (hereafter, PM selection) contains 10~569~260 objects\footnote{Note that the difference in the number of sources with the ones in MC21 comes from the different cut in radius, now being of $15^{\circ}$ instead of $20^{\circ}$.}.

\subsection{Neural Network classifier}
\label{sec:NN}

In order to improve the separation of the MW foreground from the LMC stars, we use classifiers exploiting all the information available in the \gaianospace~DR3 catalogue. Starting from a reference sample where both types of objects are labelled, we train a classifier that uses the DR3 data to optimize the separation. Then, we apply the trained classifier to our base dataset and check its performance with several validation subsets. This is an approach already used in other works; for instance, \citet{schmidt22} applies a Support Vector Machine classifier trained on a sample where the MW-LMC distinction is based on StarHorse \citep{anders22} distances. However, they apply it to data from both \gaianospace~EDR3 and the Visible and Infrared Survey Telescope for Astronomy (VISTA) survey of the Magellanic Clouds system (VMC; \citealt{cioni11}), limiting the number of objects available. Here we use only \gaia data and therefore obtain larger samples.

\subsubsection{Description of the \gaia training sample}
\label{sec:NNtrainingsample}
The training sample is a crucial element for the performance of a classifier. It needs to have the same observational data we use (\gaianospace~DR3), needs to be as representative of the problem sample as possible and, at the same time, the classification of its elements should be very reliable. Otherwise, the trained classifier will inherit the problems of the training sample, from biases in the selection to errors in the classification. A first possible approach for building a training sample is to use real data, that is, to use a sub-sample of our base dataset which has an (external) accurate classification of its objects into MW and LMC. We have identified two possible options for this approach; on the one hand, we can use samples of RR-Lyrae and Cepheid stars. Since distances for these objects can be accurately determined using period-luminosity relations, they can be located with precision in the LMC and thus distinguished from foreground objects. However, the samples available in this case are rather small, and composed of very specific types of stars. They are not representative of our global samples, which contains star of all types. On the other hand, we can use the distances in StarHorse EDR3 \citep{anders22} to distinguish MW from LMC objects; however, these distances are based on specific priors for MW/LMC and thus impose some preconditions on the objects, with the risk of propagating these preconditions to our classification. Furthermore, StarHorse only reaches a bright limit $G\leq18.5$ and therefore does not cover our faint limit $G=20.5$, making it not representative of our problem. For these reasons we have preferred not to use these samples for the training of the classifiers, but we have used them as validation samples to check our results, as described in Section \ref{sec:NNvalidation}.

A second possible approach, the one we have adopted, is to use representative simulations. A suitable training sample for the classifier would be a simulation based on stellar populations similar to the problem ones and with simulated observations mimicking the \gaia data. As part of the mission preparation, the \gaia Object Generator (GOG) \citep{luri14} was developed, and has been regularly updated. It produces realistic simulations of the \gaia data, and specifically contains separate modules for the simulation of the MW and the LMC stellar content. We have used GOG to produce a training dataset that, like our base sample, corresponds to a simulation of a $15^\circ$ radius area around the LMC centre, defined as $(\alpha, \delta) = (81.28^\circ, -69.78^\circ)$, and the LMC simulation has been tailored to make it compatible with recent estimations of the mean distance and systemic motion obtained from EDR3 data: a distance of $49.5$ kpc \citep{pietrzynski19} and a systemic motion of $\mu_{\alpha*} = 1.858$ $\mathrm{mas}\,\mathrm{yr}^{-1}$, $\mu_{\delta} = 0.385$  $\mathrm{mas}\,\mathrm{yr}^{-1}$ as in MC21.

\begin{figure*} 
    \centering
    \includegraphics[width=1\textwidth]{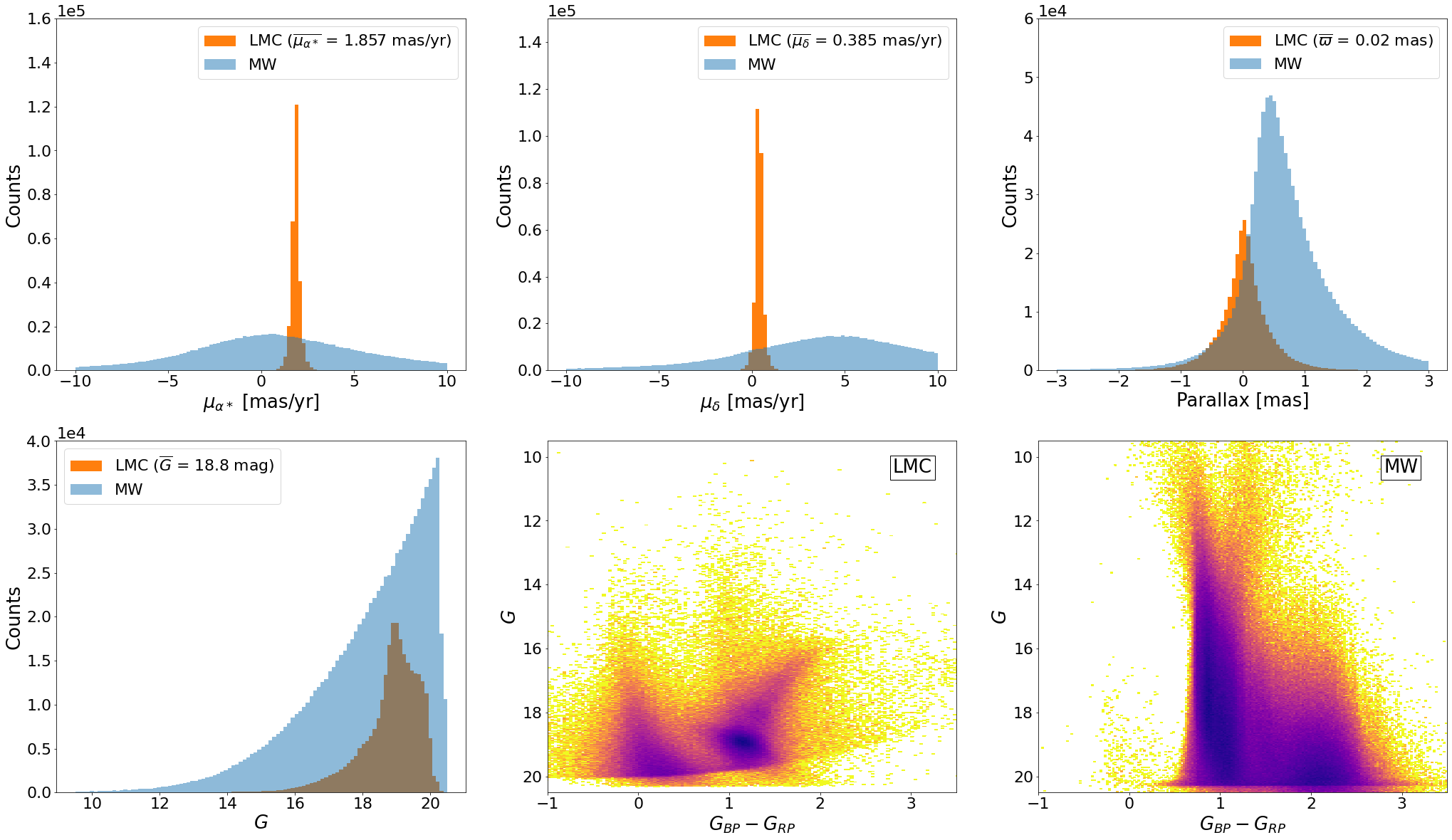}
    \caption{Characteristics of the GOG simulated samples. Top left and middle: distribution of proper motions in right ascension and declination, respectively. In orange and blue, the LMC and the MW training samples. Top right: parallax distribution. Bottom left: magnitude $G$ distribution of the simulated samples. Bottom middle and right: colour-magnitude diagram of the LMC and MW, respectively. Color represent relative stellar density with darker colors meaning higher densities.}
    \label{fig:GOG}
\end{figure*}

This \gaia training dataset is divided in two parts, one for the MW and the other for the LMC. The LMC simulation contains only 277~178 stars, a number too small when compared with real data. This is due to the design of the GOG simulator; to provide a realistic spatial distribution of the LMC simulation, it is based on a pre-defined catalogue of OGLE stars, providing real positions (see details in \citet{luri14}). The MW simulation, on the contrary, is based on a realistic galactic model, and generates a number of stars that matches the observations. This difference would give a too small LMC/MW ratio of objects, and we have corrected it by retaining only a random $20$\% fraction of the MW simulation, resulting in a total of  1~269~705 stars. Furthermore, during the trial and error phase of our selection of the configuration for the NN, we found that the classification results for the test samples (taken from the simulation data) were rather insensitive to changes in this ratio, with almost perfect ROC curves. The characteristics of the resulting simulations are summarised in Figure~\ref{fig:GOG}. 

\begin{figure*}
    \centering
    \includegraphics[width=1\textwidth]{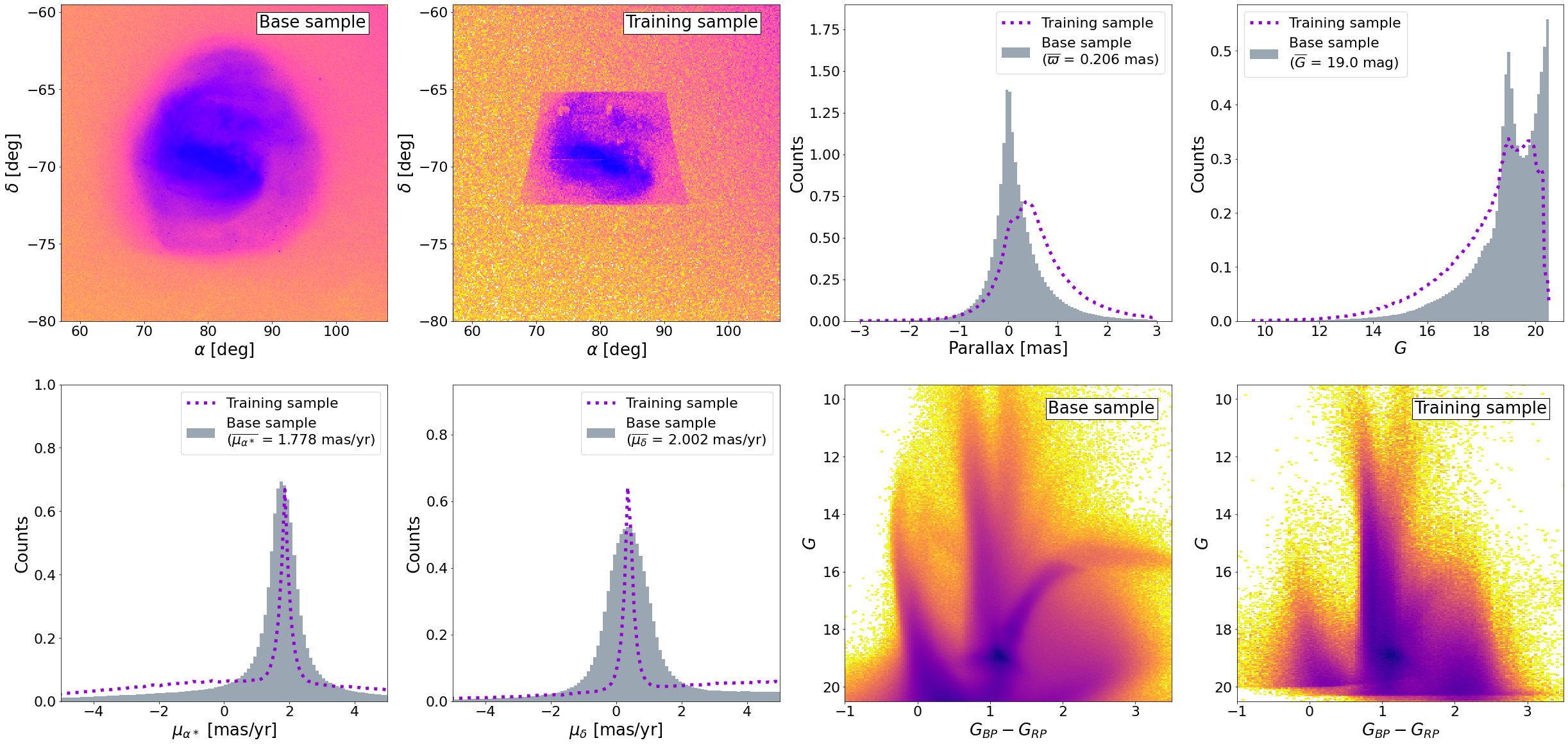}
    \caption{Comparison between the \gaia base and training samples. Top from left to right: density distribution in equatorial coordinates of the \gaia base and \gaia training samples in logarithmic scale, parallax and G-magnitude distributions. Bottom from left to right: proper motion distributions in right ascension and declination and colour-magnitude diagrams for the \gaia base and training samples. In the histograms, in gray we show the \gaia base sample, while in dotted purple we show the \gaia training sample. In the color-magnitude diagrams, color represent relative stellar density with darker colors meaning higher densities.}
    \label{fig:LMCbasesample}
\end{figure*}

The merging of these two simulations constitute our training sample, and in Figure~\ref{fig:LMCbasesample}, we compare it with the \gaia base sample. These plots show that the \gaia training sample approximately matches the main characteristics of the \gaia base sample, but its limitations are also apparent; the distribution of the LMC stars in the sky forms a kind of square, owing to its origin based on an extraction of the OGLE catalogue; the colour-magnitude diagram (CMD) for the LMC simulation is not fully representative at the faintest magnitudes, with a lack of stars and an artificial cut line; and the distributions of parallaxes and proper motions do not completely match. In spite of that, we consider these samples representative enough and we will check its performance with several validation samples to confirm its suitability. 

\subsubsection{Training the classifier }
\label{sec:NNtraining}

To implement a classifier we have used the sklearn Python module \citep{sklearn}. This module contains a variety of classifiers that can be applied to our problem, given the available \gaia data: position ($\alpha$, $\delta$), parallax and its uncertainty ($\varpi$, $\sigma_{\varpi}$), proper motion and their uncertainties ($\mu_{\alpha*}$, $\mu_\delta$, $\sigma_{\mu_{\alpha*}}$, $\sigma_{\mu_\delta}$) and \gaia photometry ($G$, $G_{BP}$, $G_{RP}$). Using the training sample described in the previous section we trained a classifier to distinguish the MW foreground objects from the LMC objects in our base sample.

\begin{figure}
    \centering
    \includegraphics[width=0.50\textwidth]{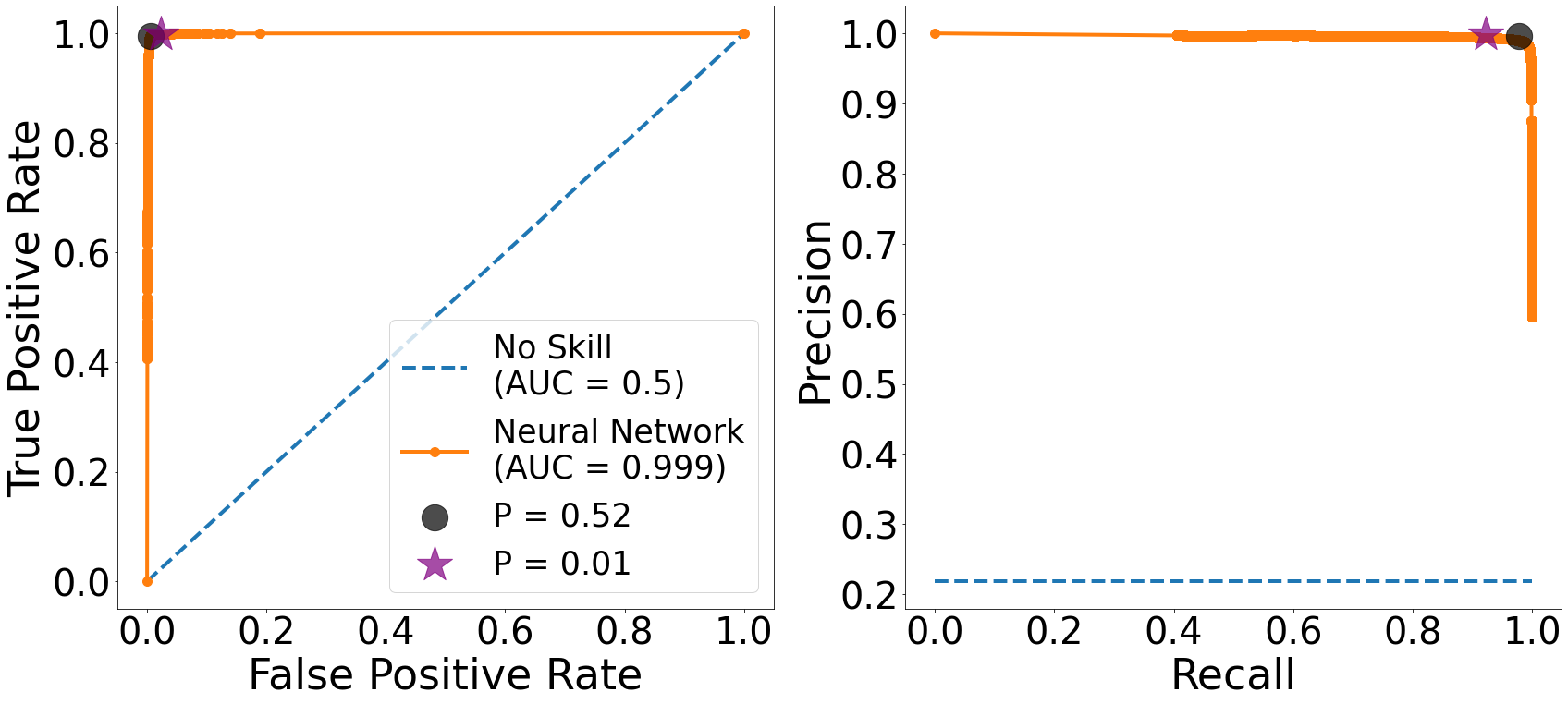}
    \caption{Evaluation metrics for the Neural Network classifier performance. Left: Receiver Operating Characteristics (ROC) curve. The black dot is in the ``elbow'' of the ROC curve and it shows the best balance between completeness and purity. The purple star shows the completeness threshold. Right: Precision-Recall curve. In both cases we compare our model (orange solid curve) with a classifier that has no class separation capacity (blue dashed curve).}
    \label{fig_roc_precission_recall}
\end{figure}

In a first stage we have tried a variety of algorithms and evaluated them internally using our simulated dataset: we have split it in two parts, $60$\% for training the algorithm and $40$\% to test its results. We have evaluated its performance by generating the corresponding Receiver Operating Characteristic (ROC) curve and calculating the Area Under the Curve (AUC). The ROC curve is one of the most important evaluation metrics for checking the performance of any classification model. It summarizes the trade-off between the true positive rate and false positive rate using different probability thresholds. The AUC of the ROC curve is another good classifier evaluator. The larger the AUC, the better the classifier works. An excellent model has AUC near to $1$ which means it has a good measure of separability. When AUC $=0.5$, it means the model has no class separation capacity. From these results we have selected three algorithms that were providing the best results: Random Forest, K Nearest-Neighbors and a Neural Network. In all three cases the ROC curve was almost perfect, similar to the one in Figure~\ref{fig_roc_precission_recall} corresponding to the Neural Network case.

After testing these three algorithms with the validation datasets described above (RR-Lyrae, Cepheids and StarHorse) and checking that they retained most of the RR-Lyrae and Cepheids when Completeness was prioritised (low probability threshold) we finally selected the Neural Network (NN) algorithm. We discarded the K Nearest-Neighbors because this type of algorithm may be too sensitive to the particularities and representativeness of the training sample (which as we have seen is limited). This was indeed the case with our samples, where for instance the square-like shape of the training sample was clearly showing in the classification results for the base sample. We also discarded the Random Forest algorithm because it produced a less sharp MW/LMC distinction, thus finally retaining the NN classifier.

Focusing on the NN classifier, we tested a few configurations and settled in a NN with 11 input neurons, corresponding to the 11 \gaia parameters listed above; three-hidden-layers with six, three and two nodes, respectively; and a single output which gives for each object the probability $P$ of being a LMC star (or, conversely, the probability of not being a MW star). A $P$ value close to $1$ ($0$) means that the object is highly likely to be of the LMC (MW). We notice that a wider exploration of NN configurations is possible, and we could test selection priorities other than Purity or Completeness (see below) in the classification, but we leave this exploration for a future work. We used the Rectified Linear Unit (ReLU) as the activation function. Our model optimizes the log-loss function using stochastic gradient descent with a constant learning rate. The L2 regularization term strength is 1e-5.\footnote{Readers interested in using the Neural Network developed in the paper can contact the corresponding author.}

In the left panel of Figure~\ref{fig_roc_precission_recall}, we show the Receiver Operating Characteristics (ROC) curve of our NN classifier. We obtain an AUC equal to 0.999 which means that our classifier separates with high-precision the LMC and MW stars in the (simulated) test sample. In the right panel of Figure~\ref{fig_roc_precission_recall}, we show the Precision-Recall curve. It is another useful metric to evaluate the classifier output quality when the classes are very imbalanced. Precision (ratio of true positive vs. total of stars classified as LMC) is a measure of result relevancy, while recall (ratio of true positives vs. total LMC stars) is a measure of how many truly relevant results are returned. As for the ROC curve, it shows the trade-off between precision and recall for different probability thresholds. 

A final warning regarding the performance of our NN. Both the ROC (AUC) and the Precision-Recall curve show an almost perfect classifier, but these results correspond to its application to the fraction of our simulated sample used for testing. In the next section, and in order to evaluate the performance with real data, we will check the NN results when applied to real samples with independent classifications. 

\subsubsection{Applying the classifier to the \gaia base data}
\label{sec:NNapplication}
\begin{figure}
    \centering
    \includegraphics[width=0.50\textwidth]{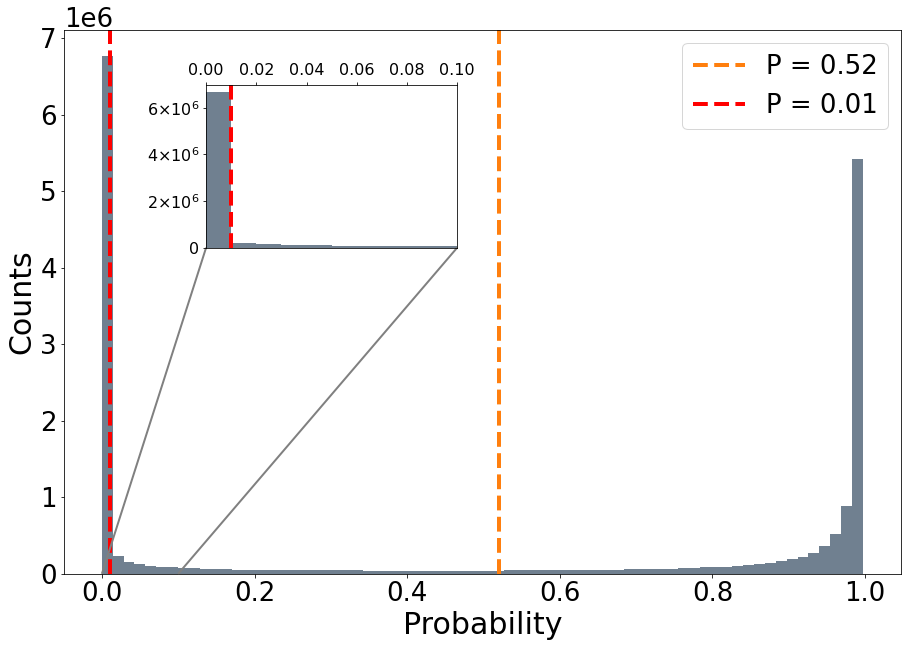}
    \caption{Probability distribution of the \gaia base sample for the Neural Network classifier. A Probability value close to 1 (0) means a high probability of being a LMC (MW) star.}
    \label{fig_proba}
\end{figure}

Once the NN is trained, we apply it to the \gaia base sample and obtain probabilities for each of its objects\footnote{The classification probability of each object will be made available in electronic form at the CDS via anonymous ftp to cdsarc.u-strasbg.fr (130.79.128.5) or via http://cdsweb.u-strasbg.fr/cgi-bin/qcat?J/A+A/}. The resulting probability distribution is shown in Figure \ref{fig_proba}. We can notice two clear peaks, one with probability close to 0 and another with probability close to 1. These peaks correspond to stars that the classifier identifies clearly as MW and LMC, respectively. In between there is a flat tail of intermediate probabilities.

To obtain a classification using the probabilities generated by the classifier for each star, we need to fix a probability threshold $P_{cut}$. If $P>P_{cut}$, the star is considered to belong to the LMC; if $P<P_{cut}$, the star is considered to belong to the MW (alternatively, we could leave stars with intermediate probabilities as unclassified). By fixing a low probability threshold we seek not to miss any LMC object, and the resulting LMC-classified sample to be more complete at the price of including more "mistaken" MW stars. On the contrary, by fixing a high probability threshold, we can make the resulting LMC-classified sample to be purer (less mistakes), at the price of missing some LMC stars and thus obtaining a less complete sample. 

The purity-completeness trade-off is a decision that will define the properties of the resulting sample and therefore can have an effect on the results obtained from it. In this work we have defined three different samples to explore the effects of this trade-off: 

\begin{enumerate}
    \item Complete sample ($P_{cut} = 0.01$). In this case a cut at small probabilities prioritizes completeness, making sure that no LMC objects are missed at the prize of an increased MW contamination. The cut value was chosen by inspecting the probability histogram of the classification (Figure \ref{fig_proba}), and selecting the limit of the main peak of small probability values.
    
    \item Optimal sample ($P_{cut}=0.52$). In this case the probability cut was chosen to be optimal in a classification sense; the value corresponds to the ``elbow'' of the ROC curve (Figure \ref{fig_roc_precission_recall}), which is in principle the best balance between completeness and purity. 
    
    \item Truncated Optimal sample ($P_{cut}=0.52$) plus an additional cut for $G > 19.5$ mag.
\end{enumerate}

We introduced the third case because, after examining the results for the Optimal sample, we noticed that the faint tail of its magnitude distribution most likely corresponds to MW stars; MW stars exponentially increase at fainter magnitudes, while LMC stars quickly decrease after $G \simeq 19.5$ (see discussion in the next section). Furthermore, with this cut we avoid a region in the faint end where the LMC training sample is not representative, as discussed above; removing these stars can reduce the MW contamination (see Section \ref{sec:NNvalidation}) and also discards the stars with larger uncertainties, and therefore less useful for our kinematic analysis. A further selection could be made by excluding regions of the CMD diagram where contamination is more likely, but given the cleanliness of the LMC diagrams in Figure~\ref{fig_histogrames_classificador} we deemed this not necessary. 

Finally, for each of the four samples we consider two datasets. First, the full sample where we assume that all the stars have no line-of-sight velocity information. Second, a sub-sample of the first one where we only keep stars with \gaianospace~DR3 line-of-sight velocities. We refer to these sub-samples as the corresponding $V_{los}$ sub-samples. The number of stars per dataset is in the second and third column of Table \ref{table_astrometry}, respectively, together with the mean astrometric information.

\subsubsection{Comparison of classifications}
\label{sec:NNcomp}
\begin{figure}
    \centering
    \includegraphics[width=0.50\textwidth]{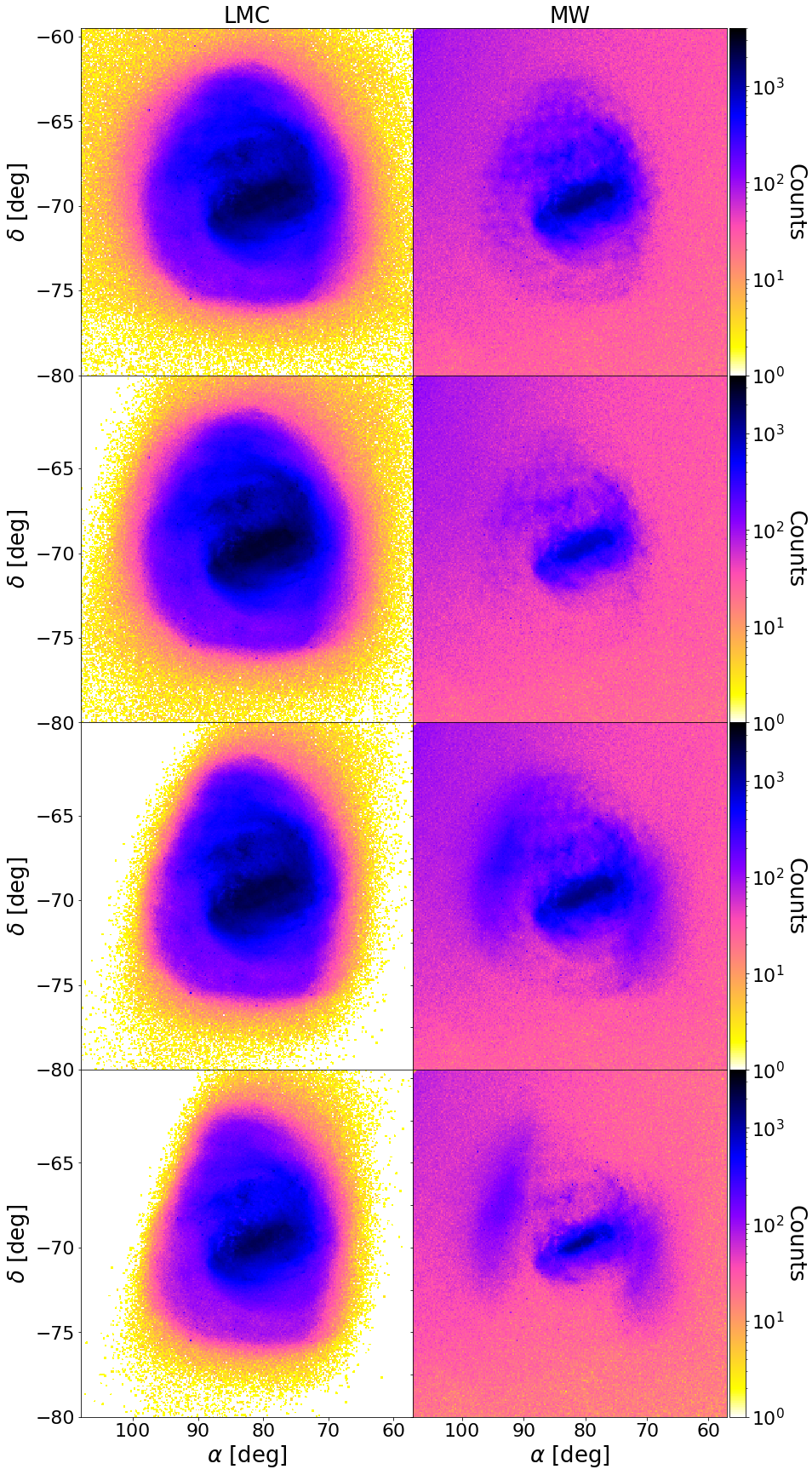}
    \caption{Sky density distribution in equatorial coordinates of both the LMC (left) and MW (right) sample obtained from the different classifiers. First row: proper motion selection classification. Second row: Complete NN classification. Third row: Optimal NN classification. Fourth row: Truncated Optimal NN classification. Note that in the fourth row we are doing a cut in magnitude $G > 19.5$ for both the LMC and MW samples, and therefore the total number of stars is reduced.}
    \label{fig_comparison_spatial}
\end{figure}

The sky density distributions for the classified LMC/MW members in our different samples are shown in Figure~\ref{fig_comparison_spatial}. 
In the left column, we show the LMC selection in each of the samples, while in the right column, we show the sources classified as MW. Each row corresponds to one selection strategy: proper motion selection (first row) followed by the three NN based ones. As expected, the results of the proper motion based selection are very similar to that described in MC21.  

We note here that the limited spatial distribution of the LMC training sample, the square region in top-left panel of Figure \ref{fig:LMCbasesample}, does not pose a problem for extrapolating the membership beyond this region, since an anomalous classification in the LMC outskirts is not observed in these figures. In order to evaluate the extrapolation performance, we have also tested the NN classifier when not taking into account the positional information; the results show that even in this extreme case the classifier does not have problems with the spatial distribution of the resulting samples.

We also note that sources classified as belonging to the MW by all four samples show an overdensity in the most crowded region of the LMC, that is, the bar, indicating misclassifications of LMC stars. We also see that, as expected by the definition of the probability cut, the more complete the LMC sample, the less stars are classified as belonging to the MW. In this respect, a cross-matching of the proper motion selection sample and the Complete sample shows that the second almost completely contains the former: of the 10~569~260 stars of the Proper motion sample, 10~432~704 of them are included in the Complete sample, and the Complete sample contains almost two million additional stars.

We also see that the dispersion of the astrometric parameters diminish from the NN Complete to the NN Truncated Optimal samples. This is expected, since the stricter sequence of selection criteria lead to a higher similarity in distance and velocity inside the samples. 

\begin{table*}
\centering
\begin{tabular}{|l|c|c|c|c|c|c|c|c|}
\hline
LMC sample                &  N  & $N_{vlos}$ & $\overline{\varpi}$ & $\sigma_\varpi$ &   $\overline{\mu_{\alpha*}}$ & $\sigma_{\mu_{\alpha*}}$ & $\overline{\mu_{\delta}}$ & $\sigma_{\mu_{\delta}}$ \\ \hline
Proper motion selection & 10 569 260 & 29 678 & -0.006        & 0.333          & 1.800                & 0.408                   & 0.369               & 0.541                  \\
NN Complete      & 12 116 762    & 30 749 & -0.008            & 0.382             & 1.808                   & 0.563                      & 0.348                  & 0.653                     \\
NN Optimal     & 9 810 031    & 22 686 & -0.016            & 0.346             & 1.819                   & 0.446                      & 0.364                  & 0.488                     \\
NN Truncated Optimal     & 6 110 232    & 22 686 & -0.008            & 0.211             & 1.820                   & 0.353                      & 0.357                  & 0.423                     \\\hline
\end{tabular}
\caption{Comparison of the LMC samples number of sources and mean astrometry between the proper motion selection (MC21) and the Neural Networks. Parallax is in $\mathrm{mas}$ and proper motions in $\mathrm{mas}\,\mathrm{yr}^{-1}$.}
\label{table_astrometry}
\end{table*}

\begin{figure*}
    \centering
    \includegraphics[width=1\textwidth]{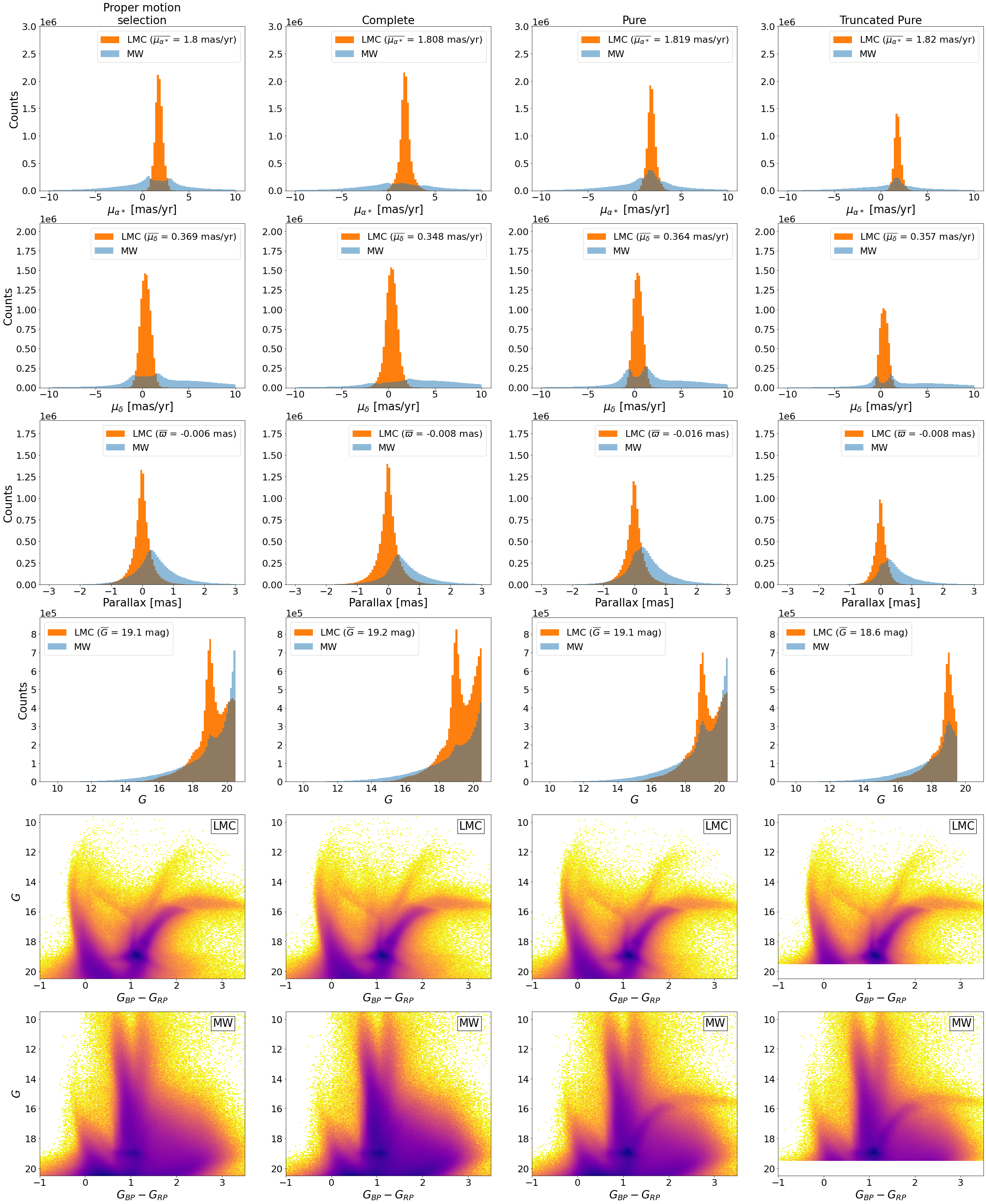}
    \caption{Astrometric and photometric characteristics of the LMC and MW samples. From left to right: PM sample, NN Complete, NN Optimal and NN Truncated Optimal samples. In the first four rows, we show distributions of proper motion in right ascension and declination, parallax and G magnitude, respectively, of the LMC (orange) and MW (blue) samples. In the last two rows, we show the colour-magnitude diagram of the samples classified as LMC and MW, respectively. Color represents the relative stellar density, with darker colors meaning higher densities.}
    \label{fig_histogrames_classificador}
\end{figure*}

In Figure \ref{fig_histogrames_classificador} we compare the astrometry and photometry distribution of the different LMC samples. In the proper motion selection sample, we see that the sharp cut in proper motion imposed makes the distribution of proper motion to be narrow around the bulk motion of the LMC, while in the MW classification, two small peaks are present, following a continuation of the LMC peak. Clearly, some LMC stars are misclassified as MW using the sharp cut in proper motion. This misclassification is not visible in the NN Complete sample and is present again in the more restrictive Optimal and Truncated Optimal samples. The parallax distribution in the four LMC samples are very similar, being the Truncated Optimal sample the most narrow. The $G$ magnitude distributions are quite different in the four LMC selections. Both the PM and the NN samples show a peak in $G$ magnitude around $G\sim 19$ mag, which corresponds to the LMC sample, and a secondary peak at the limiting magnitude of $G=20.5$, corresponding to MW contamination. For this reason, as described above, we define the Truncated Optimal sample by removing the secondary peak in the Optimal sample. Reversely, the MW selection in all cases should show an exponential distribution in $G$, though the PM, the Complete and the Optimal samples show a secondary peak of different significance amongst them around $G \sim 19$\,mag. The CMD of all LMC samples is very similar. Small differences only appear in the MW selection of the Optimal and Truncated Optimal sample which contain, as expected, sources of the red giant branch of the LMC, which the NN classifier misclassifies as MW. 

\subsubsection{External validation of the classification}
\label{sec:NNvalidation}
\begin{figure}
    \centering
    \includegraphics[width=0.50\textwidth]{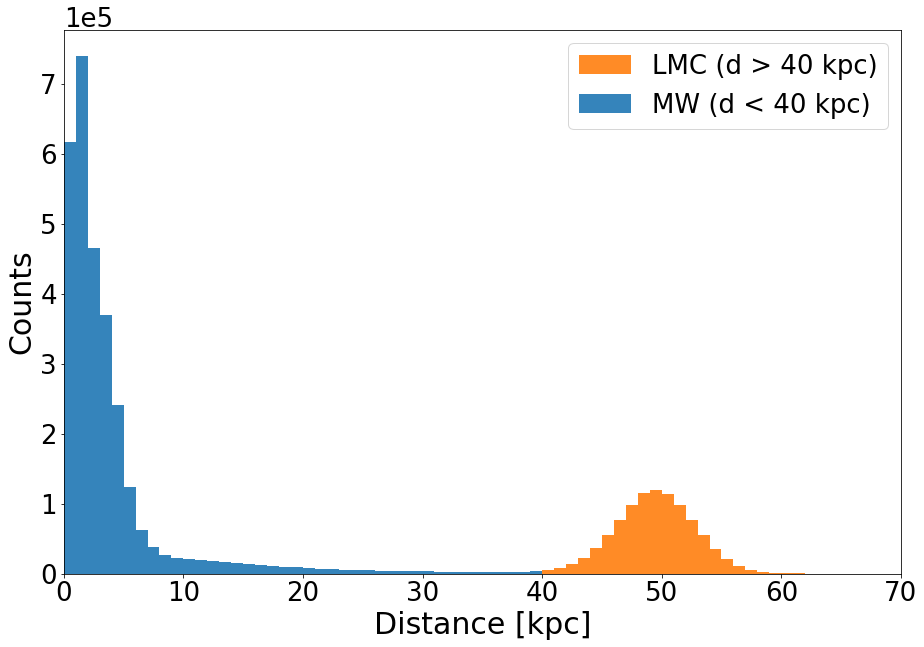}
    \caption{Distance distribution of the StarHorse validation sample. In blue (orange), the StarHorse stars classified as MW (LMC) according to the $d = 40$ kpc criteria.}
    \label{fig_SH}
\end{figure}

As indicated in previous sections, to validate the results of our selection criteria we compare them with external independent classifications. To do so, we have cross-matched our base sample with three external samples:

\begin{itemize}
    \item LMC Cepheids \citep{ripepi22}: we used the paper's sample of 4~500 Cepheids as a set of high-reliability LMC objects. To obtain the \gaia EDR3 data we cross-matched the positions given in the paper with the \gaia EDR3 catalogue, using a 0.3" search radius to obtain high confidence matches, thus retaining 4~485 stars. Finally, we introduced a cut of 15$^{\circ}$ radius around the LMC center (mimicking our base sample), leading to a final selection of 4~467 LMC Cepheids. 
    \item LMC RR-Lyrae \citep{cusano21}: similarly to the above, we used the paper's sample of 22~088 RR-Lyrae as high-reliability LMC objects. After the cross-match with the \gaia EDR3 catalogue, the sample is reduced down to 22~006 stars and after the in 15$^{\circ}$ radius cut around the LMC center we obtain a final sample of 21~271 LMC RR-Lyrae. 
    \item StarHorse \citep{anders22}: we cross-matched this catalogue with the \gaia EDR3 data using a cut of 15$^\circ$ radius around the LMC center and obtained a sample of 3~925~455 stars. Following a similar criteria to the one proposed in \citep{Schmidt20,schmidt22}, we have separated MW and LMC stars through the StarHorse distances, but making a cut at $d = 40$ kpc. This decision is motivated by the distance distribution of the StarHorse sample, which is shown in Figure~\ref{fig_SH}. A cut in $d = 40$ kpc gives a very restrictive classification, minimizing the contamination of MW stars (see discussion below). We thus obtain a StarHorse LMC sample with 985~173 stars and a StarHorse MW sample with 2~940~282 stars. Notice that being based on StarHorse, this sample contains stars only up to $G=18.5$.
\end{itemize}

The Cepheids and RR-Lyrae samples contain objects classified with high reliability as LMC stars, so they serve as a check of the completeness of our classification for LMC objects (``how many we lose''). On the other hand, the StarHorse sample is helpful to estimate the contamination caused by wrongly classified MW stars, although this can only be taken as an indication, since the StarHorse classification itself is not perfect. Furthermore, given the very stringent criteria used for the separation in StarHorse (the cut in $d = 40$ kpc), the resulting estimation of MW contamination in the classification will be a ``worst case''. 

In Table~\ref{table_validation} we summarize the comparison of the results of our four classification criteria applied to stars contained in the three validation samples. It can be seen that the completeness of the resulting LMC classifications is quite good, usually above $85$\%, as shown for the results with the Cepheids, RR-Lyrae and StarHorse LMC validation samples. The exception is the Truncated Optimal sample, where the completeness is reduced for the RR-Lyrae, due to the cut in faint stars.

\begin{table*}
\centering
\begin{tabular}{|l|c|c|c|c|}
\hline
Stars classified as LMC         & \begin{tabular}[c]{@{}c@{}}LMC Cepheids\\ (4~467)\end{tabular} & \begin{tabular}[c]{@{}c@{}}LMC RR-Lyrae\\ (21~271)\end{tabular} & \begin{tabular}[c]{@{}c@{}}LMC StarHorse\\ (985~173)\end{tabular} & \begin{tabular}[c]{@{}c@{}}MW StarHorse\\ (2~940~282)\end{tabular} \\ \hline
Proper motion selection         & 4 366 (97.7\%)                                                   & 18 673 (87.8\%)                                                   & 970 173 (98.5\%)                                                    & 704 932 (24.0\%)                                                   \\
NN Complete       & 4 407 (98.7\%)                                                   & 20 223 (95.1\%)                                                   & 970 719 (98.5\%)                                                    & 722 750 (24.6\%)                                                   \\
NN Optimal           & 4 160 (93.1\%)                                                   & 17 860 (84.0\%)                                                   & 832 733 (84.5\%)                                                    & 627 619 (21.3\%)                                                   \\
NN Truncated Optimal & 4 160 (93.1\%)                                                   & 14 750 (69.3\%)                                                   & 832 733 (84.5\%)                                                    & 627 619 (21.3\%)                                                   \\ \hline
\end{tabular}
\caption{Matches of the classified LMC members in our four considered samples against the validation samples. Percentages are calculated with respect to the total number of stars given below the sample name.}
\label{table_validation}
\end{table*}

\begin{figure}
    \centering
    \includegraphics[width=0.50\textwidth]{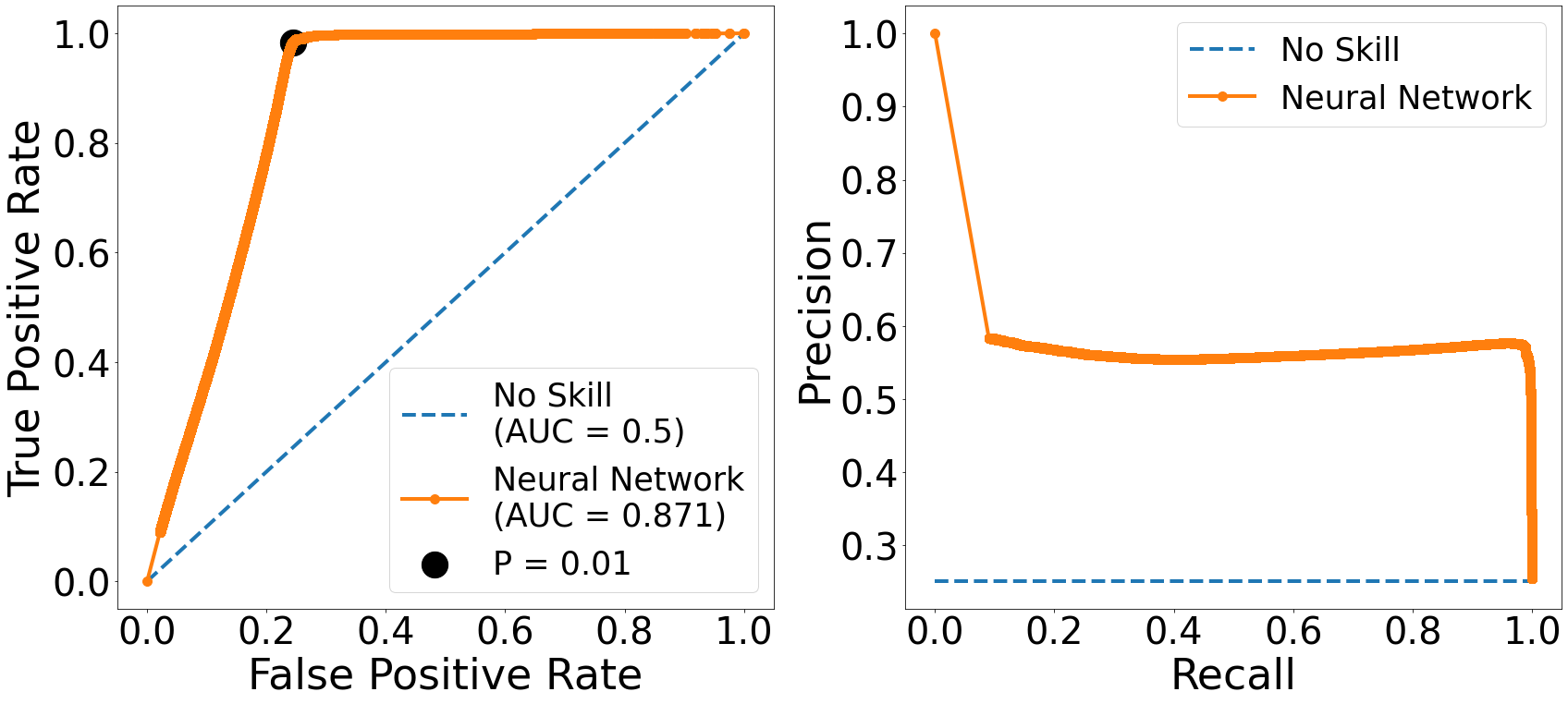}
    \caption{Same as Fig.~\ref{fig_roc_precission_recall} for the Neural Network classifier performance using the StarHorse sample.}
    \label{fig_roc_precission_recall_SH}
\end{figure}

On the other hand, the relative contamination by MW stars in the samples is more difficult to assess. We have to rely on the StarHorse distance-based classification as an external comparison, with the caveat that this classification contains its own classification errors. To do so, we re-calculate the Precision-Recall curve, but this time taking the StarHorse classification as a reference; the result is shown in Figure \ref{fig_roc_precission_recall_SH}. We can see that the Precision remains quite flat for almost all the range of the plot, that is, for all the range of probability threshold values. This indicates that the relative contamination (percentage of stars in the samples that are MW stars wrongly classified as LMC stars) is similar in the Complete and Optimal samples (the more restrictive we are, the more MW stars we remove, but also we lose more LMC stars). Taking the Precision values in Figure \ref{fig_roc_precission_recall_SH} indicates that, using the classification based on SH distances as a reference, the relative contamination of our samples could be around 40\%; this is a worst case, since we have used a very restrictive distance cut ($40\,$kpc), and using less restrictive cuts (down to $10\,$kpc) the estimation of the contamination lowers to $\sim30$\%. These numbers have to be taken with care, since the MW-LMC separation based on the SH distances is not perfect, just another possible classification criteria that in fact is using less information than our criteria. As pointed in the SH paper: {\it These populations are now clearly visible as overdensities in the maps, although a considerable amount of stars still has median distances that fall in between the Magellanic Clouds and the MW - a result of the multimodal posterior distance distributions} \citep{anders22}. 

\begin{figure*}
    \centering
    \includegraphics[width=1\textwidth]{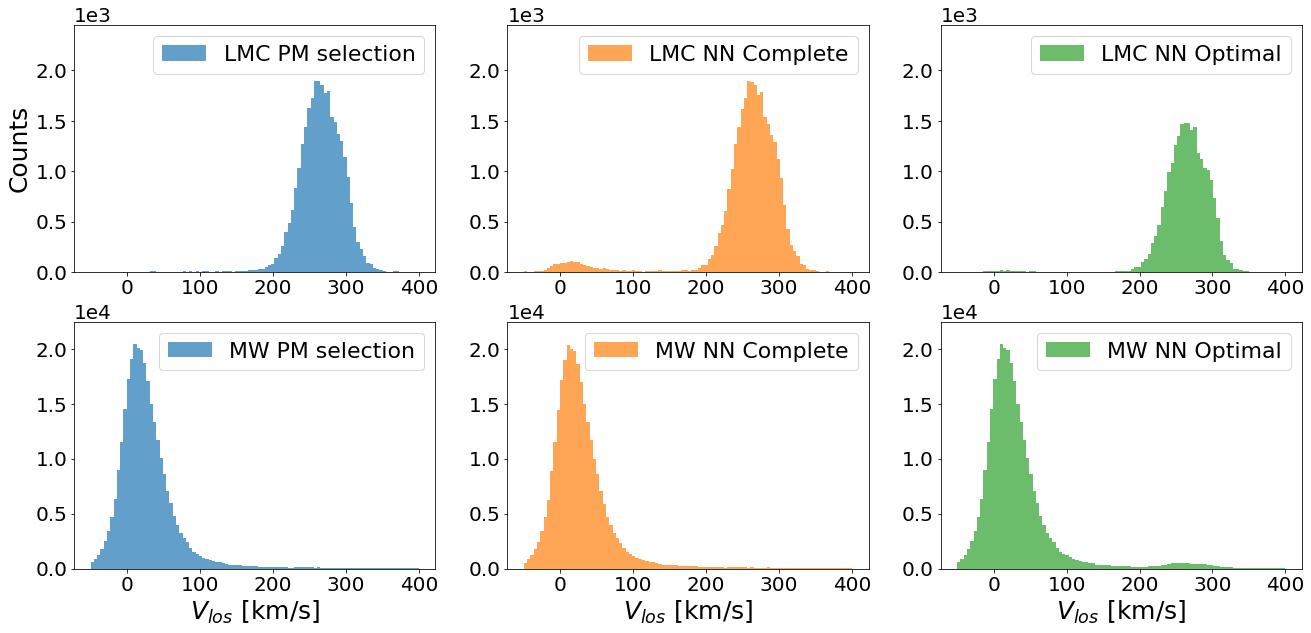}
    \caption{Line-of-sight velocity distribution for the stars classified as LMC (top) and MW (bottom). We show the three $V_{los}$ sub-samples of the PM selection (left), NN Complete (middle) and NN Optimal (right) samples.}
    \label{fig:vlos_histogram}
\end{figure*}

These results point out to a possible contamination by MW stars in our samples around some tens of percentage but we can do an additional check using the line-of-sight velocities in \gaianospace~DR3, which are available only for a (small) subset of the total sample. These line-of-sight velocities are not used by any of our classification criteria and have different mean values for the MW and LMC (therefore providing an independent check). In Figure \ref{fig:vlos_histogram} we have plotted the histograms of line-of-sight velocities separately for stars classified as MW and LMC, and it is clear from them that the contamination of the LMC sample is reduced, probably significantly below the levels suggested above. For instance, if we consider the LMC NN Complete sample and (roughly) separate the MW stars with a cut at $V_{los} < 125$ \kms, we estimate the MW contamination to be around $5\%$. However, since the subset of \gaianospace~DR3 stars with measured line-of-sight velocities contains only stars at the bright end of the sample ($G \lesssim 16$), this check is not fully representative either.

Finally, we made a new query to the Gaia archive similar to that defined in Sect. 2.1. This time, we made a selection from the \textit{gaia\_source} table in \gaia DR3 with a 15$^\circ$ radius in a nearby region with homogeneous sky density. This way we can make an estimation of the MW stars expected in a regions similar to that covered by our \gaia base sample. From this new query we obtain $4~240~771$ stars, so we would expect a similar number of MW stars in the region we selected around the LMC. Given that the Gaia base sample contains $18~783~272$ objects and the number of objects classified as LMC (Table \ref{table_astrometry}) is around $6-12$ million, the number of stars classified as MW is around $12-6$ million; therefore, we can conclude that our NN LMC samples prioritise purity to completeness since there are too many stars classified as MW (an excess of $2$ to $8$ million). This is also evident from the right panels of Figure \ref{fig_comparison_spatial} where the distribution of stars classified as MW shows the pattern of LMC contamination.

\section{Coordinate transformations and validation}
\label{sec:trans}

\subsection{Coordinate transformations}

Since the main goal of this work is to look at the internal kinematics of the LMC, we review the coordinate transformations used to compute the LMC-centric velocities. To do so, we revisit the formalism introduced in \citet{vdm01} and \citet{vdm02} and describe the two-step process used to transform the \gaia heliocentric measurements to the LMC reference frame (full details are given in Appendix \ref{appendix_vdM}). 

\begin{figure}
    \centering
    \includegraphics[width=0.50\textwidth]{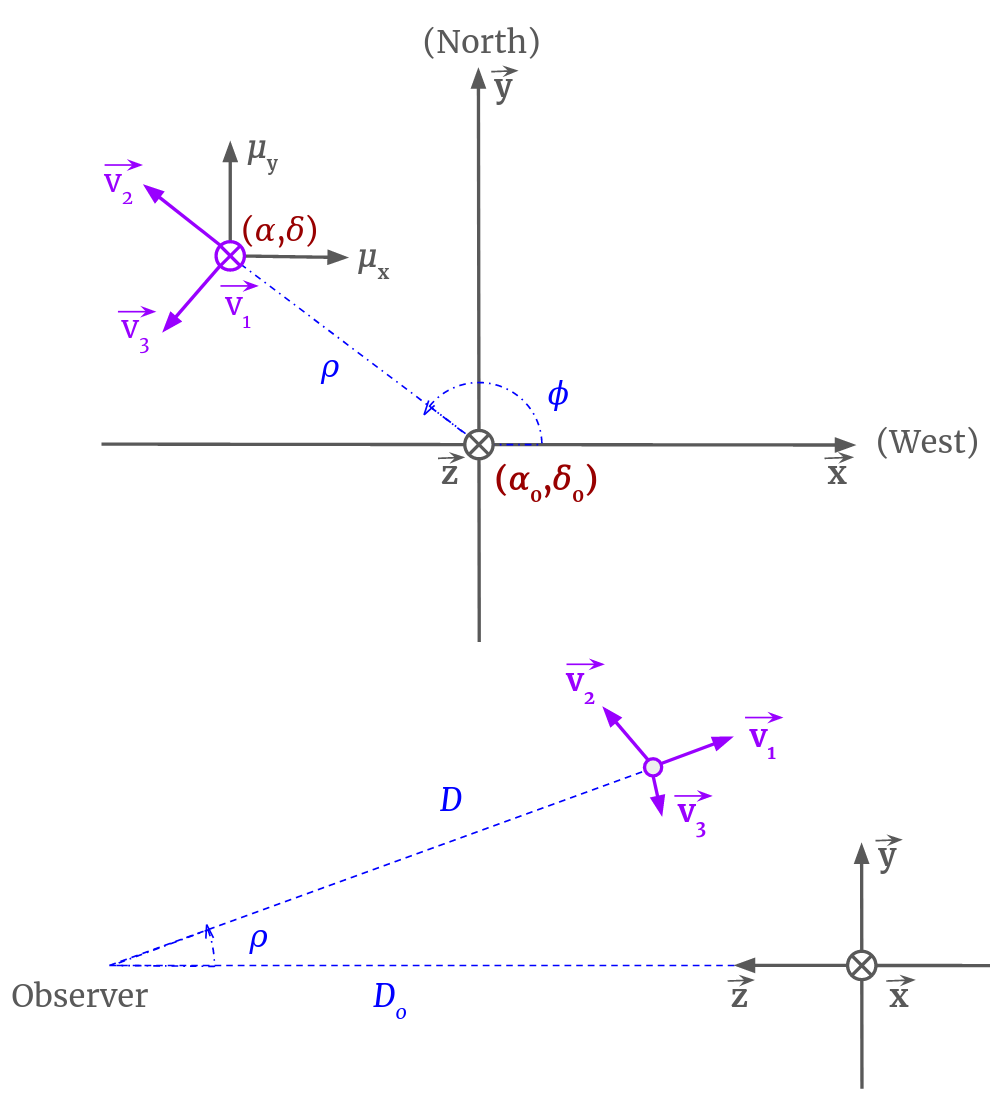}
    \caption{Schematic view of the observer-galaxy system. The LMC center $O(\alpha_0, \delta_0, D_0)$ is chosen to be the origin $O$ of the $(x, y, z)$ coordinate system. Top: Projected view of the sky. All vectors and angles lie in the plane of the paper. The angles $\rho$ and $\phi$ define the projected position on the sky of a given point with coordinates ($\alpha, \delta$). Bottom: Side view of the observer-galaxy system. The distance from the observer to the LMC center is $D_0$ and the distance from the observer to an object is $D$. The component $v_1$ lies along the line-of-sight and points away from the observer.}
    \label{Fig1_schema}
\end{figure}

First, we introduce a Cartesian coordinate system $(x,y,z)$, whose origin, $O$ is placed at $(\alpha_0,\delta_0,D_0)$, the LMC centre. The orientation of the $x$-axis is anti-parallel to the right ascension axis, the $y$-axis parallel to the declination axis, and the $z$-axis towards the observer. This is somehow similar to considering the orthographic projection -- a method of representing three-dimensional objects where the object is viewed along parallel lines that are perpendicular to the plane of the drawing -- of the usual celestial coordinates and proper motions. See Figure~\ref{Fig1_schema} for a schematic view of the observer-galaxy system. We refer to this reference frame as the orthographic projection centred at the LMC.

Second, we transform from the $(x,y,z)$ frame to the final Cartesian coordinate system whose reference plane is the LMC plane, $(x',y',z')$. It consists of the superposition of a counterclockwise rotation around the $z$-axis by an angle $\theta$, followed by a clockwise rotation around the new $x'$-axis by an angle $i$. With this definition, the $(x',y')$ plane is inclined with respect to the sky tangent plane by an angle $i$. Face-on (face-off) viewing corresponds to $i=0^\circ$ ($i=90^\circ$). The angle $\theta$ is the position angle of the line-of-nodes or, in other words, the intersection of the $(x',y')$-plane and the $(x,y)$-plane of the sky. By definition, it is measured counterclockwise from the $x$-axis. In practice, $i$ and $\theta$ will be chosen such that the $(x',y')$-plane coincides with the plane of the LMC disk. Therefore, we refer to this final reference frame as the LMC in-plane reference system.

\begin{figure}
    \centering
    \includegraphics[width=0.45\textwidth]{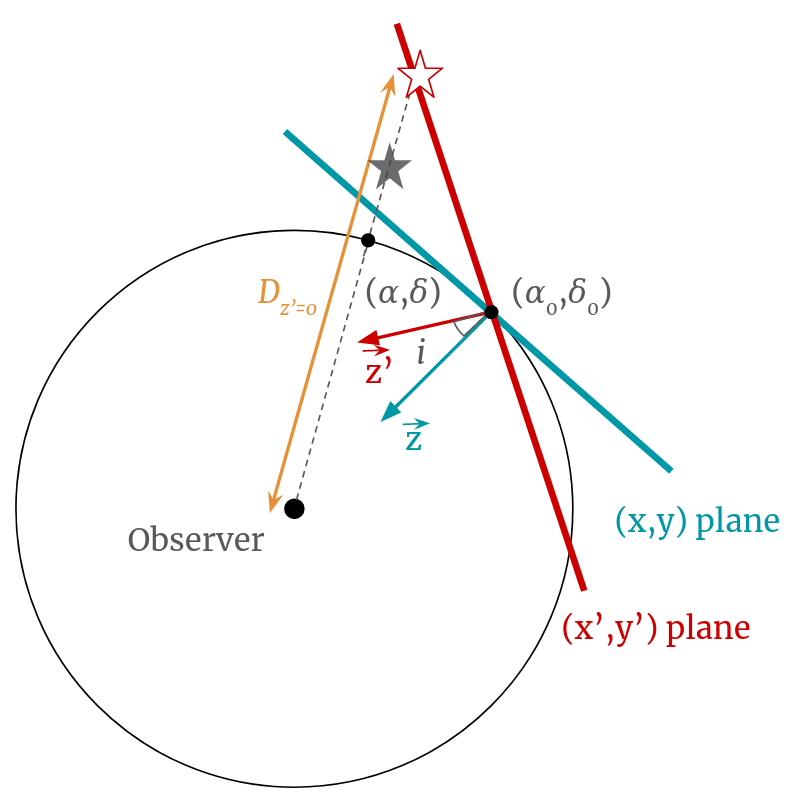}
    \caption{Schematic representation of the reference frames used, all of them centred on the LMC centre $(\alpha_0,\delta_0)$. In blue, we show the orthographic reference frame, $(x,y,z)$, while in red, we show the Cartesian LMC frame, $(x',y',z')$. We also show the position of a real star (solid dark gray), its projection in the LMC cartesian frame under the imposition of $z'=0$ (red frame).}
    \label{fig:schema_z=0}
\end{figure}

Since we do not have reliable information for individual distances because the parallaxes are very small and close to the noise \citep[MC21, ][]{edr3_astrometric}, we assume that all the stars lie on the LMC disc plane, as an approximation. Thus, we impose $z'$ to be zero, which leads to a distance of $D_{z'=0}$ (different to the real one) for each star. In Figure~\ref{fig:schema_z=0} we show a schematic representation of what this assumption implies. We represent the position of a real star in dark gray, while the white star in red solid line is the projection of the real star on the LMC plane. With this strategy all LMC stars are assumed to lie on its plane.

When these two rotations are applied, and the stars are made to lie in the LMC plane, for each star we have LMC-centric positions $(x', y', 0)$ and velocities $(v_{x'}, v_{y'}, v_{z'})$. The last step is to make these velocities internal by removing the LMC systemic motion (see details in Appendix \ref{append_sys_motion}). In this work, as in MC21, we consider the following LMC parameters: 
$i = 34^\circ$, $\theta = 220^\circ$, $(\alpha_0, \delta_0) = (81.28^\circ, -69.78^\circ)$, and $(\mu_{x,0}, \mu_{y,0}, \mu_{z,0})$ = $(-1.858, 0.385, -1.115)$ $\mathrm{mas}\,\mathrm{yr}^{-1}$, where we have taken into account that our $x$ and $z$-axes have the opposite sense from the one considered in MC21. These values are derived assuming a specific centre, the same one as we use in this work. The distance to the LMC centre is assumed to be $D_0 = 49.5$ kpc \citep{pietrzynski19}. 

As shown in Eq. (\ref{eq_vlos}), the formalism presented in \citet{vdm01} and \citet{vdm02} allows taking into account line-of-sight velocities, which is something that could not be done when using MC21 transformations. As detailed in Section \ref{sec:NNapplication}, in this work we deal with two different datasets: the full samples, without line-of-sight velocity information and the sub-samples of stars with individual line-of-sight velocities. For the former, we estimate each star line-of-sight velocity by taking into account its position and proper motion and the global parameters of the LMC plane (full details in Appendix \ref{append_no_vlos}). 

\begin{figure}
    \centering
    \includegraphics[width=0.50\textwidth]{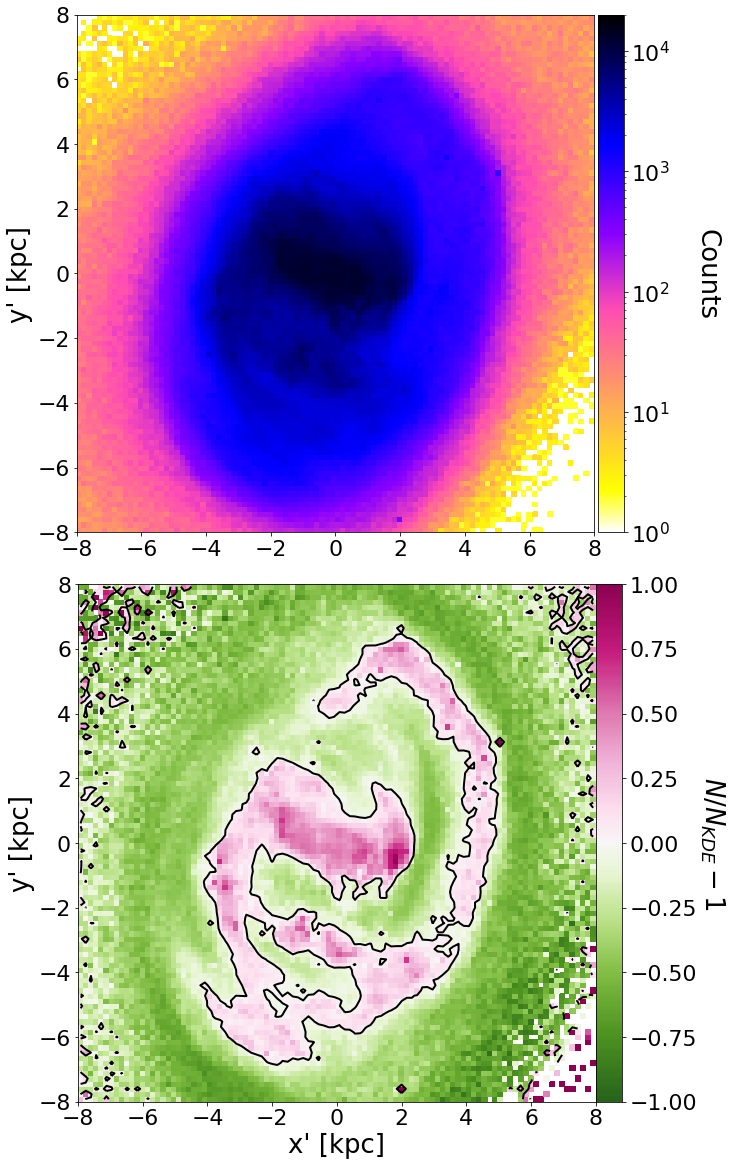}
    \caption{Top: LMC density map for the NN Complete sample. Bottom: LMC overdensity map for a 0.4kpc-bandwidth KDE. A black line splitting the overdensities from the underdensities is plotted. Both maps are shown in the $(x',y')$ Cartesian coordinate system.}
    \label{fig_LMC_density}
\end{figure}

In the top panel of Figure~\ref{fig_LMC_density}, we show the LMC density map for the NN Complete sample in the LMC cartesian coordinate system. The density maps for the rest of the three samples are analogous and show the same morphological features as in the corresponding sample of the left column of Figure~\ref{fig_comparison_spatial}. Here we want to point out that the coordinate transformation from the heliocentric equatorial system to the LMC cartesian system inverts the vertical axis, so now the spiral arm starts at negative $x'$ and $y'$, and the deprojection of the inclination angle in the sky makes the galaxy elongated along the vertical axis.

\subsection{Validation of the formalism with a N-body simulation}
\label{sec:simu}
\begin{figure*}[!h]
     \centering
     \includegraphics[width=0.39\textwidth]{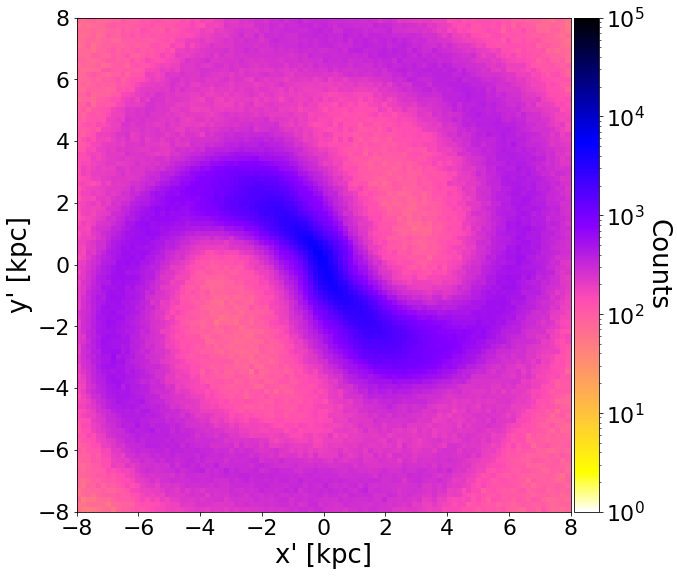}
     \includegraphics[width=0.5\textwidth]{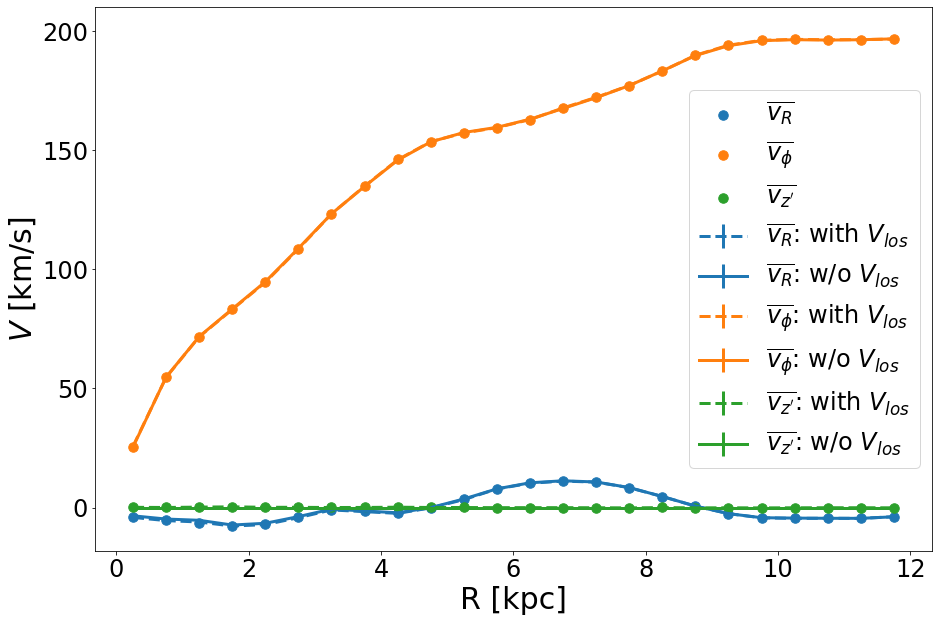}\\
     \includegraphics[width=0.95\textwidth]{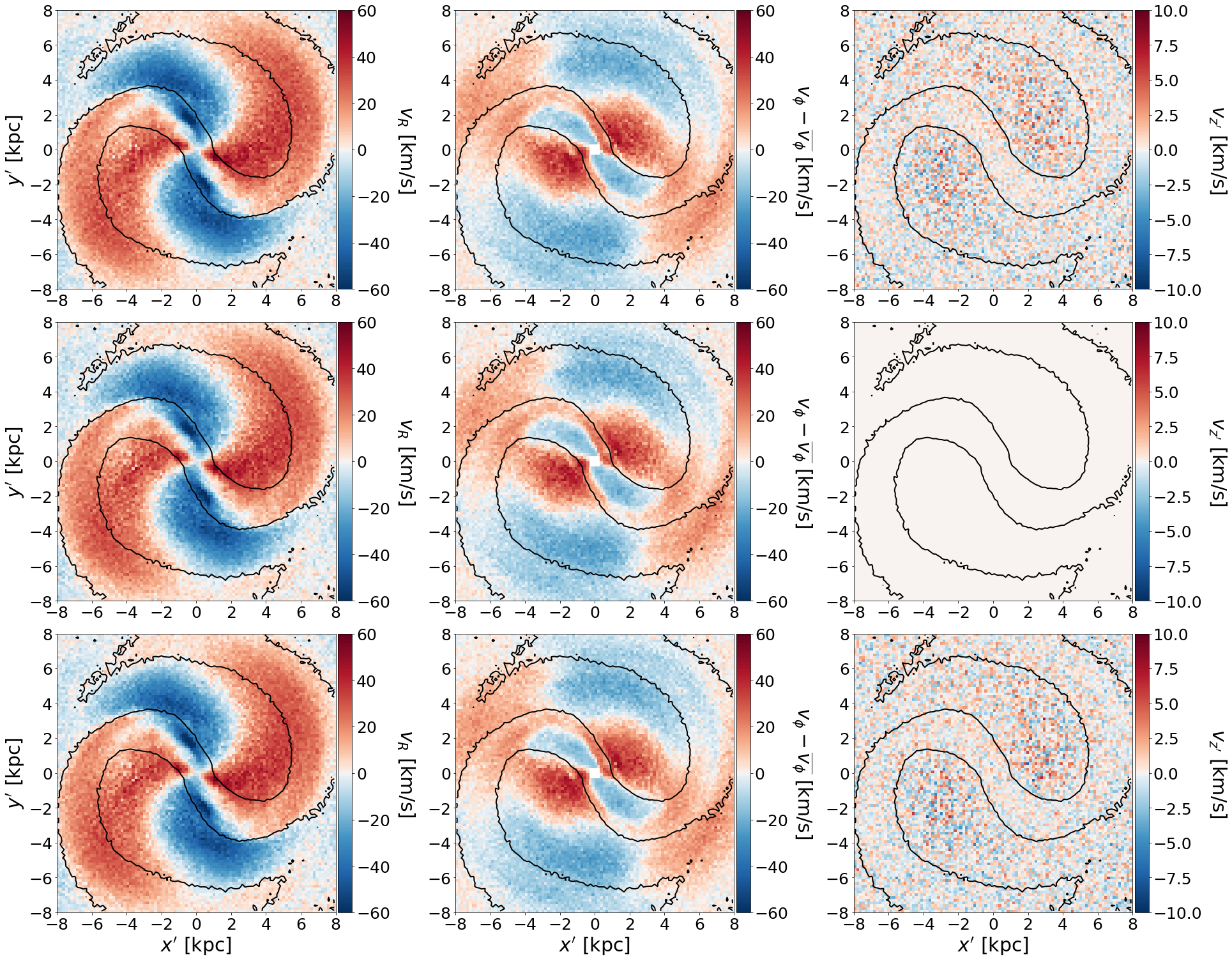}
     \caption{Simulation B5. First row left panel: density distribution in logarithmic scale. First row right panel: velocity profiles of the B5 simulation. In blue, orange and green, for the radial, tangential and vertical components, respectively, when taking into account the full velocity information (dashed lines), or when $V_{los}$ is not available (solid lines). Differences are negligible and both curves overlap. Scatter points show the real velocity profiles. Second row: N-body simulation maps (left, radial; middle, residual tangential; right, vertical), without applying any coordinate transformation. Third row: same as above computed applying Section \ref{sec:trans} coordinate transformations without line-of-sight information. Fourth row: same as above computed applying Section \ref{sec:trans} coordinate transformations with line-of-sight information. The black line shows the contour corresponding to overdensity equal to zero for a 0.4kpc-bandwidth KDE (see details in Section~\ref{sec:overdensity})}
     \label{fig:simu_velocity_maps}
 \end{figure*}

In this section we use N-body simulations to test the new formalism introduced above. We use the "B5" isolated barred galaxy simulation of \citet{rocafabrega2013}, which consists of a live disc of $5$ million particles and a Toomre parameter of $Q=1.2$, and a live NFW halo. The disc to halo mass ratio is the appropriate so that the simulation develops a strong bar and two spiral arms which are transient in time. The snapshot used in this analysis corresponds to an evolution time of $T\simeq 500\,$Myr, and the density distribution is shown in the top left panel of Figure~\ref{fig:simu_velocity_maps}. Using the 6D information of positions and velocities at this given time, we make the following exercise. First, we convert the galactocentric cartesian coordinates to heliocentric equatorial $(\alpha,\delta,d,\mu_{\alpha^*},\mu_{\delta},V_{los})$ at the line-of-sight, spatial orientation, distance and systemic motion of the LMC. Then, we consider these particles as a data set and we apply the same formalism described in Section~\ref{sec:trans} to compute the coordinates in the LMC frame and the velocities in cylindrical coordinates. The radial component (left panels) indicates the motion towards/outwards from the galactic centre, the residual tangential velocity (middle panels) is obtained by subtracting the tangential velocity curve to the tangential velocity component, and indicates the motion with respect to the tangential curve. The vertical component (right panels) indicates the motion across the galactic plane. 

We apply the coordinate transformations twice. In the first case, we impose, as in the LMC full samples, that $V_{los}$ is not available and use the internally derived from Eq.~\ref{eq:v1_int}. Secondly, we use the available $V_{los}$, as in the sub-samples, from the 3D velocity data. We compute velocity profiles and velocity maps with the simulation data as follows. The same procedure is performed when applying it to the LMC samples in Section~\ref{sec:velocity}. Each curve is obtained by computing the median value of all stars located in radial bins of $0.5$kpc-width in the $(x',y')$-plane. The error in each bin is computed as the division between the median absolute deviation and the square-root of the number of stars. The resulting velocity profiles are shown in the top right panel of Fig.~\ref{fig:simu_velocity_maps}. The velocity maps are obtained by plotting the median in $100\times 100$ bins in the $(x', y')$-plane from $-8\,$kpc to $8\,$kpc. The resulting velocity maps are shown in the second to fourth row panels of Fig.~\ref{fig:simu_velocity_maps}. In the second row we show the radial, residual tangential and vertical velocity maps obtained directly from the N-body simulation. In the third and fourth rows, we show the velocity maps when $V_{los}$ is not available and when it is, respectively.

From the velocity profiles and maps, we note that the $V_{los}$ approximation used, when this is not available, does not modify the velocity profiles as seen in the top right panel of Fig.~\ref{fig:simu_velocity_maps}, nor the radial and residual tangential velocity maps (see left and middle panels of Fig.~\ref{fig:simu_velocity_maps}). The only effect is in the vertical component in the case $V_{los}$ is not available. In this case, we obtain $V_z'=0\,$\kms, which is a consequence of the fact that the internal line-of-sight velocity is estimated by computing the derivative of the distance as function of time (see Eq.~\ref{eq:v1_int}), which makes $V_z'$ to become null when substituting into the analogous Eq.~\ref{eq_vlos} for the internal motion. In the other two cases, namely when we use the input data or the derived $V_z'$ from the $V_{los}$ we obtain a median profile and median velocity map centered at zero within the Poisson noise.
In the radial and residual tangential velocity component (left and middle panels) we clearly see the quadrupole effect due to the presence of a rotating bar. As expected, the change in sign in $V_R$ occurs along the major and minor axes of the bar and the residual tangential velocity is minimum along the bar major axis. Also, we successfully validated the coordinate transformation formalism by artificially inflating the vertical component ten times larger than that of the original B5 simulation to make the radial, tangential and vertical velocity components comparable in range.

In conclusion, the formalism used to derive the velocities in the LMC frame, when the line-of-sight velocity is not available does not introduce any bias in the velocity profiles or maps. The most important assumption in the formalism is that all stars lie on a plane.

\section{Analysis of the velocity profile and maps}
\label{sec:velocity}
In this section we analyse the velocity profiles (Section~\ref{sec:velprofiles}) and velocity maps (Section~\ref{sec:velmaps}) for the four samples, and their corresponding $V_{los}$ sub-samples, described in Section~\ref{sec:data}. To allow comparison between density and kinematics, we overplot in the velocity maps the overdensity contour, as described in Section~\ref{sec:overdensity}. 

\subsection{LMC velocity profiles}
\label{sec:velprofiles}

Here we analyse the velocity profiles in the LMC coordinate system. We compute the LMC velocity profiles by using a similar methodology to that used in MC21, and as specified in Section~\ref{sec:simu}. 

\begin{figure*}
    \centering
    \includegraphics[width=0.95\textwidth]{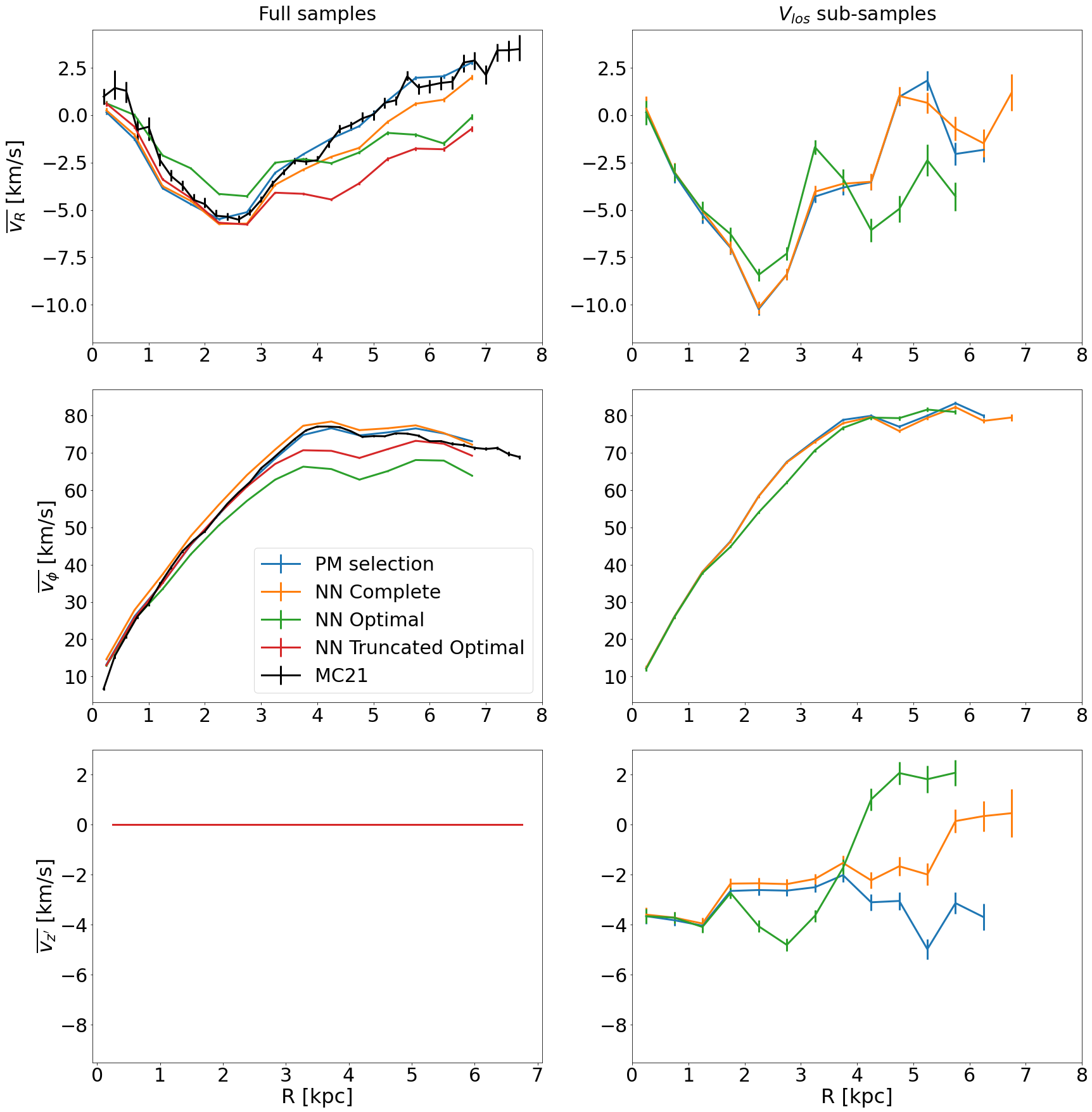}
    \caption{Velocity profiles for the four LMC samples in the case $V_{los}$ is not available (left) and when it is available (right). From top to bottom, radial, tangential and vertical velocity profile. Each curve corresponds to one LMC sample: PM selection (blue), NN Complete (orange), NN Optimal (green) and NN Truncated Optimal (red). In black, the radial and rotation curve published in MC21 (their Figure 14) is shown. For most of the bins, the error bar is small and cannot be seen. Only bins with more than 300 sources are plotted.}
    \label{fig:LMC_velo_profile}
\end{figure*}

In the left panels of Figure~\ref{fig:LMC_velo_profile}, we show the velocity profiles (radial, tangential and vertical, from top to bottom) for each of the four full LMC samples. In all samples, the radial velocity profile slightly decreases with radius up to $2.5$\,kpc, where it increases again. The tangential velocity profile shows the rotation curve of the galaxy, having a linear growth until $R\sim4\,$kpc in all samples and becoming flat in the outer disc. The maximum tangential velocity varies between LMC samples, with a maximum difference of $\sim15\,$\kms at $R=4.7\,$kpc between the NN Complete and NN Optimal samples. We use these rotation curves to derive residual tangential velocity maps in Section~\ref{sec:velmaps}. The vertical velocity profile for the four samples is completely flat and centered at $0\,$\kms, which is a consequence of not using the observational $V_{los}$ in these samples, as mentioned in Section~\ref{sec:simu}. As mentioned above, the internal line-of-sight velocity is estimated by computing the derivative of the distance as function of time (see Eq.~\ref{eq:v1_int}), which makes $V_z'$ to become null when substituting into the analogous Eq.~\ref{eq_vlos} for the internal motion, so we will not use it in the following analysis.

We can summarise the comparison between the samples in the following points:
\begin{itemize}
    \item The PM sample profiles are almost the same as the MC21 for the radial and tangential velocity curves. 
    \item The NN Complete sample profiles are very similar to the ones of the PM selection, as also seen in the density maps (see Section~\ref{sec:NNcomp}), with a maximum difference of $\sim 2$\kms at $R\gtrsim 4.5\,$kpc in the radial component.
    \item The NN Optimal and NN Truncated Optimal, while being purer samples, have larger inward streaming motion and rotate about $\sim 10-15\,$\kms more slowly than the more complete samples in the outer disc.
\end{itemize}

In the right panels of Figure~\ref{fig:LMC_velo_profile} we show the same velocity profiles, now with the sub-samples that have full 3D velocity components, this is, with $V_{los}$. The trends observed in the left panels for the radial and tangential profiles are reproduced in the right panels. We note that the radial velocity profile of the $V_{los}$ sub-samples has a larger negative amplitude at $R=2-3\,$kpc, of around $\overline{V_R}=-10\,$\kms, compared to the $\bar{V_R}=-5\,$\kms of the full samples. Also, the tangential velocity for the NN sub-samples is now a bit larger and they are all centered around $\overline{V_{\phi}}\sim 80\,$\kms. The largest difference between the main samples and the $V_{los}$ sub-samples arises in the vertical velocity component, where we observe different trends between sub-samples: 
\begin{itemize}
    \item The NN Truncated Optimal sample provides the same profile as the NN Optimal sample. This is expected since, in this case, both sub-samples are the same because line-of-sight velocities are available up to Gaia G magnitude $G<16.$, and the truncation is performed at $G=19.5$. 
    \item All sub-samples have a slightly negative vertical velocity profile. The PM sample is the one that has a flatter profile, centered around $-3\,$\kms up to $\sim 5\,$kpc. NN Complete has an increasing trend from $-4\,$\kms to positive values at $R=6\,$kpc and NN Optimal presents a wave-like pattern having a change of sign at $R\sim 4\,$kpc. These trends are very sensitive to the imposed $\mu_{z,0}$. A small shift to the vertical component of the systemic motion will translate into a shift in the $V_z'$ profile, while differences between sub-samples arise from contamination from the MW, mostly present in the outer disc (see discussion in Section~\ref{sec:discussion}).
\end{itemize}

\subsection{Determination of the LMC overdensity maps}
\label{sec:overdensity}
In this section, we introduce and describe the mask used to highlight the LMC overdensities, such as the bar and the spiral arm.

To analyse the data, we considered $100\times 100$ bins in the $(x', y')$-plane of range $-8\,$kpc to $8\,$kpc, as when constructing the velocity maps. Then, a Gaussian kernel density estimation (KDE) of $0.4\,$kpc-bandwidth is applied. To highlight the star overdensities, we compute for each bin $N/N_{KDE}-1$ where $N$ and $N_{KDE}$ are the number of stars corresponding to the data histogram and the KDE, respectively. The mask $N/N_{KDE}-1$ has positive (negative) values for overdensities (underdensities). $N_{KDE}$ is computed by integrating the KDE for the bin area. The choice of the KDE bandwidth was empirical. We perform overdensity maps for bandwidths ranging from $0.2-1\,$kpc, and we consider that a $0.4$kpc-bandwidth fulfills our objective of highlighting both the LMC bar and arm.

In the bottom panel of Fig~\ref{fig_LMC_density}, we show the LMC overdensity map in the $(x',y')$ Cartesian coordinate system. We can clearly see how both the LMC bar and spiral arm stand out as overdensities, as shown by the black contour of overdensity equal to zero. We first observe how the LMC spiral arm starts at the end of the bar around $(-3,0)$ kpc. Then, if we analyse the spiral arm following a counter-clockwise direction, the spiral arm breaks into two parts: an inner and an outer arm. Finally, both parts join further on and continue together until the spiral arm ends, performing about a full rotation around the LMC centre.

\subsection{LMC velocity maps}
\label{sec:velmaps}

\begin{figure*}
    \centering
    \includegraphics[width=0.74\textwidth]{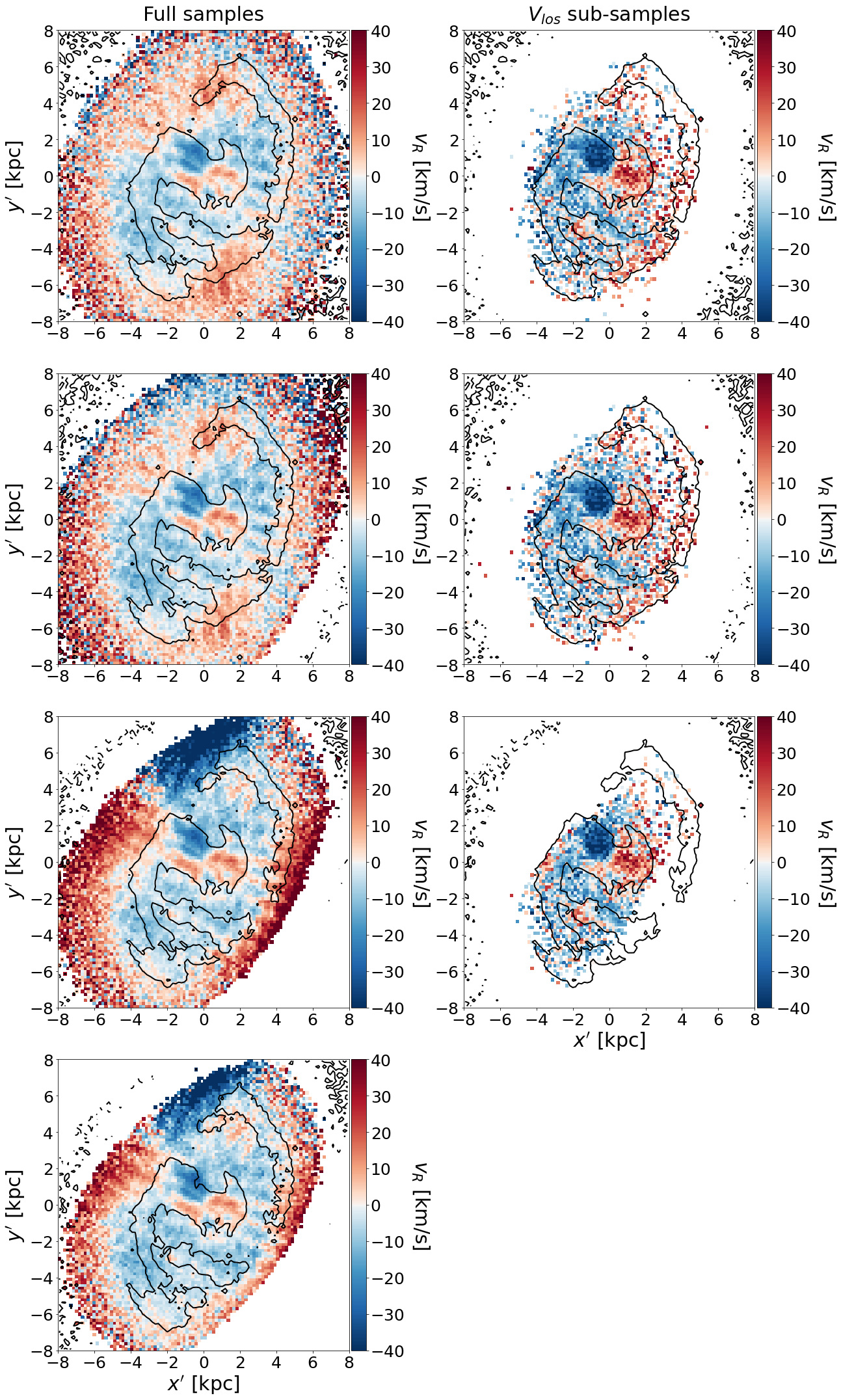}
    \caption{LMC median radial velocity maps. All maps are shown in the $(x',y')$ Cartesian coordinate system. From top to bottom: PM sample, NN Complete, NN Optimal and NN Truncated Optimal sample. Left: without line-of-sight velocity. Right: with line of sight velocity. NN Truncated Optimal $V_{los}$ sub-sample map is not shown because it is the same as the NN Optimal $V_{los}$ sub-sample (see text for details). 
    For each colormap, a black line splitting the overdensities from the underdensities for a 0.4kpc-bandwidth KDE is plotted and a minimum number of 3 (20) stars per bin is imposed when line-of-sight is (not) considered.}
    \label{fig_lmc_velocity_vr}
\end{figure*}

\begin{figure*}
    \centering
    \includegraphics[width=0.75\textwidth]{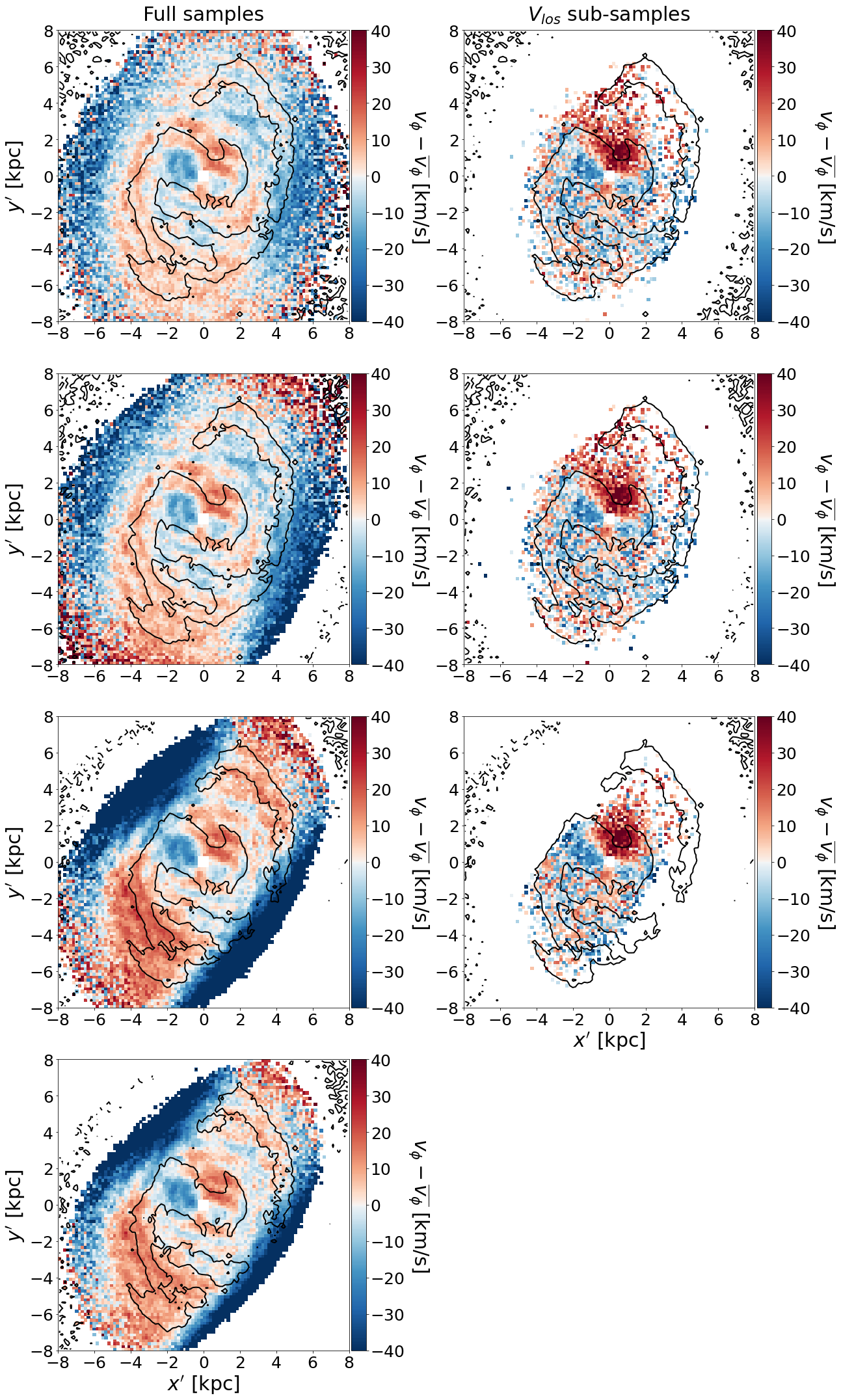}
    \caption{Same as Fig.~\ref{fig_lmc_velocity_vr} for the median residual tangential velocity.}
    \label{fig_lmc_velocity_vphi}
\end{figure*}

\begin{figure}
    \centering
    \includegraphics[width=0.39\textwidth]{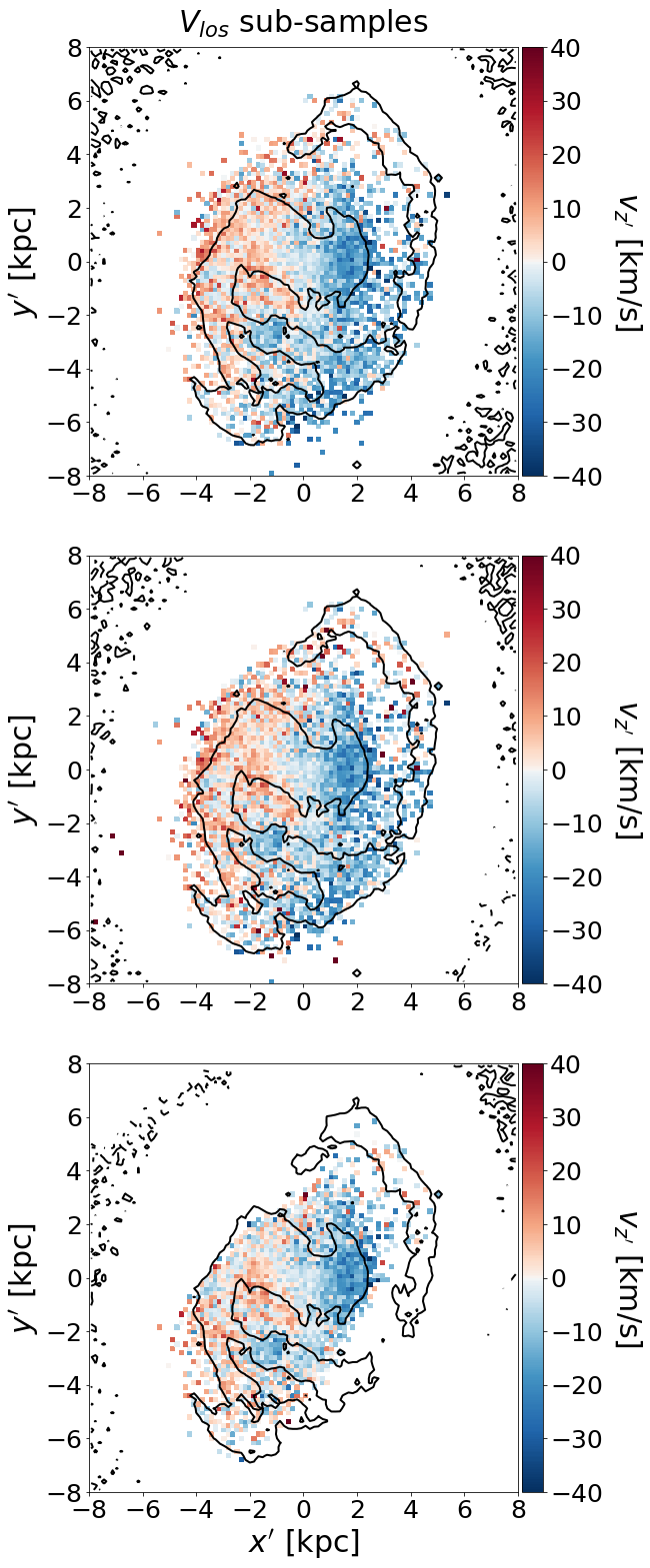}
    \caption{ Same as Fig.~\ref{fig_lmc_velocity_vr} for the median vertical velocity of the $V_{los}$ sub-samples.}
    \label{fig_lmc_velocity_vz}
\end{figure}

In this section we analyse the velocity maps in the LMC coordinate system for the four LMC samples. The results are shown in Figures~\ref{fig_lmc_velocity_vr}, \ref{fig_lmc_velocity_vphi} and \ref{fig_lmc_velocity_vz} for the radial, tangential and vertical components, respectively. Results are shown from top to bottom for PM selection, NN Complete, NN Optimal and NN Truncated Optimal samples, respectively, while left (right) panels show the velocity maps for the full ($V_{los}$) samples. The black line shows the overdensity contour corresponding the overdensity equals to zero, as in the bottom panel of Fig.~\ref{fig_LMC_density}, which helps to compare between density and kinematics.

Regarding the radial velocity maps (Fig.~\ref{fig_lmc_velocity_vr}), the quadrupole pattern already reported in MC21 related to the motion of stars in the bar is present for all samples, though for $V_{los}$ sub-samples an asymmetry clearly becomes apparent along the semi-major axis of the bar, shown by the change in sign of the radial velocity. We estimate the bar major axis is inclined with respect to the $x^{\prime}$-axis about $\sim -10$\degr. The radial velocities in the upper half of the semi-major axis have larger values (in absolute value) than those on the bottom half. Further research is required to analyse whether this asymmetry is an effect of the inclination of the bar with respect to the galactic plane, or it is an artifact of the assumption that all stars lie in the plane, though this assumption is also present in the full sample, where the asymmetry is also present but less clear. The trend in the outer disc is similar in all full samples, though a strong inward/outward motion in the outskirts of the sample is present in the NN Optimal and NN truncated optimal samples. In particular, the strong inward motion detected in the upper periphery of the NN Optimal and NN truncated samples is coherent with the region where the Magellanic Bridge connects to the LMC, and could represent in-falling stellar content from the SMC. Along the LMC spiral arm, there is a negative (inward) motion along the spiral arm overdensity when this is still attached to the bar, regardless of the sample and the number of velocity components used (left panels of Fig.~\ref{fig_lmc_velocity_vr}). After the break, there is not a clear trend. In the right panels, the $V_{los}$ sub-samples do not have enough number of stars on the spiral arms to provide a clear conclusion.

Regarding the residual tangential velocity maps (Fig.~\ref{fig_lmc_velocity_vphi}), the conclusions in the bar region are analogous to that of the radial velocity maps, namely the quadrupole pattern expected for the motion of the stars in elliptical bar orbits is present. The asymmetry in terms of larger velocity in absolute value above the bar major axis is clear in both the full samples and the $V_{los}$ sub-samples. This asymmetry in the velocity seems slightly larger in the NN Optimal and NN Truncated Optimal samples, with a maximum difference of $10\,$\kms. Along the spiral arm the residual tangential velocity is in general positive in all samples, this is, stars on the spiral arm move faster than the mean motion at the same radius, except for the piece of the arm with a density break. When comparing between samples, it represents the effect of the contamination of MW stars in the velocity maps, we see the decrease of the residual tangential velocity in the edges of the sample in the NN Optimal and Truncated Optimal samples, which could be a bias of the sample. 
Regarding the $V_{los}$ sub-samples (right panels of Fig.~\ref{fig_lmc_velocity_vphi}), there is not a clear sign of the residual tangential velocity along the piece of the spiral arm attached to the bar. 

Finally, we show in Fig.~\ref{fig_lmc_velocity_vz} the vertical velocity component for the $V_{los}$ sub-samples. We see, as in the vertical velocity profile, that the vertical velocity map has second order differences between the different samples. More complete samples, such as PM and NN Complete samples show a bimodal trend, where half of the galaxy ($x'<0$) is moving upwards while the other half ($x'>0$) is moving downwards. There also seems to be a positive gradient (in absolute value) of increasing vertical velocities from the inner to the outer disc. This could be associated to an overestimation of the disc inclination angle, or to the presence of a galactic warp \citep[e.g.][]{Choi2018}, or to the contamination of MW stars. Purer samples, such as the NN Optimal or NN Truncated Optimal still show a similar wave-like motion that can be associated to the warp or to the fact that the LMC is still not in dynamical equilibrium \citep[e.g.][]{Choi2022}. Also, there seems to be a clear negative motion of stars located at the end of the bar with $x^{\prime}>0$, which could be an evidence of the inclination of the bar with respect to the galactic plane.

\section{Biases and different evolutionary phases}
\label{sec:discussion}

Velocity maps may be affected by the choice of the galaxy parameters, namely the inclination $i$ and position angle $\theta$, or the systemic motion $(\mu_{x,0},\mu_{y,0},\mu_{z,0})$. There remains a large uncertainty in the literature on what the inclination angle of the galactic plane with respect to the line-of-sight and the line-of-nodes of the position angle are, with differences as large as $10$\degr \citep[e.g.][]{vandermarel&kallivayalil2014, Haschke&Grebel&Duffau2012,ripepi22}. In this work we use $i=34$\degr and $\theta=220$\degr, as nominal values, as in MC21 work. In order to study the possible systematic that a different inclination angle or position angle can induce in the velocity maps, we reproduce the velocity maps for the NN Complete sample and corresponding $V_{los}$ sub-sample, by varying the nominal values by $\pm 10$\degr\,, only one at a time. In general, the effect of having a larger (smaller) inclination angle, elongates (stretches) the velocity map, in either of the velocity components, and a different position angle, rotates the velocity maps. In detail, they also introduce the following trends:
\begin{itemize}
\item Regarding the inclination angle:
\begin{itemize}
    \item The variation of the inclination angle from $i-10$\degr to $i+10$\degr, can reverse the radial motion in the outer disc from being positive to negative along the $y'$ axis.
    \item The median vertical component can even change sign and become negative if using a smaller inclination angle, while if this is $10$\degr larger, a clear bi-symmetry is introduced.
    \item The inner disc is not affected, nor the residual tangential component. 
\end{itemize} 
\item Regarding the position angle, systematics are of second order and mainly affect the azimuthal angle in the disc where the motion is inward/outward and upward/downward.
\end{itemize}

\begin{figure}
    \centering
    \includegraphics[width=0.47\textwidth]{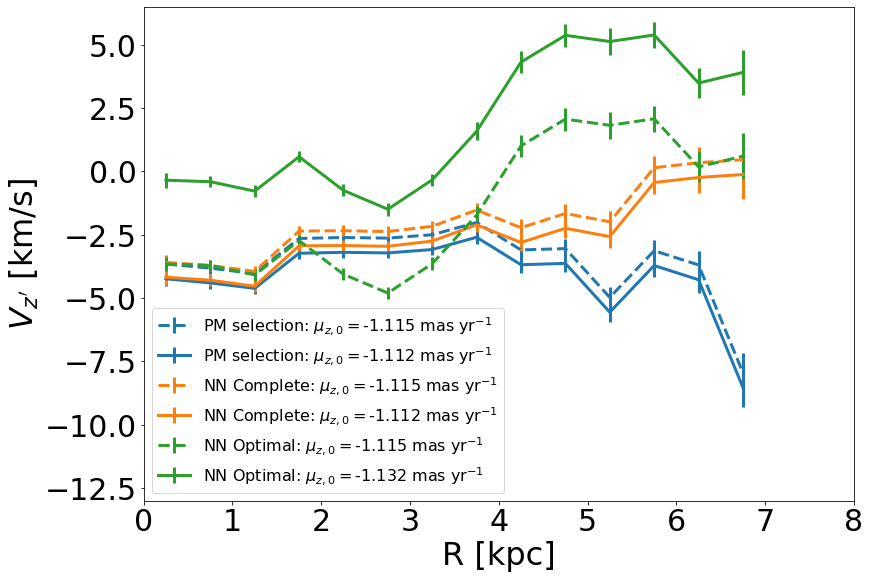}
    \caption{Stellar vertical velocity profiles of the LMC $V_{los}$ sub-samples (blue, orange and green for the PM, NN Complete and NN Optimal, respectively) for different input values of $\mu_{z,0}$: the \gaia~DR3 derived in solid lines and the MC21 adopted value in dashed lines (values given in the legend).}
    \label{fig:LMC_velo_profile_muz0}
\end{figure}

The choice of the systemic motion $(\mu_{x,0},\mu_{y,0},\mu_{z,0})$ used to compute the internal velocities in the LMC reference frame may also introduce systematics in the velocity profile and maps. In this work we have adopted the same systemic motion as in MC21 to allow a direct comparison. The availability of line-of-sight velocities in \gaia~DR3 allows a better estimation of $\mu_{z,0}$. For each $V_{los}$ sub-sample, we fit a Kernel Density Estimation of $2\,$\kms bandwidth to the distribution of line-of-sight velocities (see top panels of Fig.~\ref{fig:vlos_histogram}) and obtain at which line-of-sight velocity this distribution is maximum $V_{los,0}$. We assume, then, that the line-of-sight systemic motion, $\mu_{z,0}$ of the LMC is given by $\mu_{z,0}=V_{los,0}/D_{0}$. Results are shown in Table~\ref{table_muz0}, where we show for each sub-sample  $V_{los,0}$ and the corresponding $\mu_{z,0}$. We note that for the PM and NN Complete samples $\mu_{z,0}=-1.112\,\mathrm{mas}\,\mathrm{yr}^{-1}$ and similar to the value adopted in this work, which is of $-1.115\,\mathrm{mas}\,\mathrm{yr}^{-1}$. As discussed previously, NN Complete sample and PM sample have a very similar LMC classification and thus provide similar density and kinematic distributions. On the other hand, the line-of-sight systemic motion for the NN Optimal sample gives $\mu_{z,0}=-1.132\,\mathrm{mas}\,\mathrm{yr}^{-1}$, which can provide a different velocity profile in the vertical in-plane component, while it barely has no effect on the planar components. In Fig.~\ref{fig:LMC_velo_profile_muz0}, we show the vertical velocity profile for each of the three sub-samples when we use either the MC21 adopted value for $\mu_{z,0}$ (dashed lines) or the derived using the \gaia~DR3 line-of-sight velocities (solid lines). Note that a bias in the $\mu_{z,0}$ translates in a shift in the $V_z^{\prime}$ profile. The small difference between the PM and NN Complete samples with respect to the MC21 value falls within the error bars. For the NN Optimal sample the currently derived value shifts the median vertical velocity to be centered at zero in the inner $3\,$kpc while it is slightly oscillating towards positive values in the outer disc.

\begin{table}
\centering
\begin{tabular}{|l|c|c|}
\hline
$V_{los}$ sub-sample                &  $V_{los,0}$\, [\kms]  & $\mu_{z,0}$\, [$\mathrm{mas}\,\mathrm{yr}^{-1}$] \\ 
\hline
PM             & 260.86     & -1.112  \\
NN Complete    & 260.86     & -1.112  \\
NN Optimal     & 265.66     & -1.132  \\
\hline
\end{tabular}
\caption{Determination of the line-of-sight systemic motion. Second column gives the $V_{los}$ for which the KDE (see text for details) is maximum. Third column provides the corresponding value of $\mu_{z,0}$.}
\label{table_muz0}
\end{table}

Regarding the tangential systemic motion, $(\mu_{x,0},\mu_{y,0})$, \citet{vdm02} determines $(\mu_{x,0},\mu_{y,0})=(-1.68,0.34)\,$ $\mathrm{mas}\,\mathrm{yr}^{-1}$ using Carbon stars, while \citet{schmidt22} find $(\mu_{x,0},\mu_{y,0})=(-1.95,0.43)\,$ $\mathrm{mas}\,\mathrm{yr}^{-1}$. In this work, we use the values derived in MC21, namely $(\mu_{x,0},\mu_{y,0})=(-1.858,0.385)\,\mathrm{mas}\,\mathrm{yr}^{-1}$. The choice of $(\mu_{x,0},\mu_{y,0})$ affects the three components of the internal velocities (see Equations~\ref{eq:v1_int} and \ref{eq:v123_int}). We test how a possible change of the systemic motion within values given by different models in MC21 (their Table~5) and in the literature affects the velocity profile and maps. Regarding the MC21 values, the velocity maps do not change qualitatively, and the largest change is in the vertical velocity profile with a shift of the order of $2\,$\kms within the uncertainty range of $0.02\,\mathrm{mas}\,\mathrm{yr}^{-1}$ in either of the tangential systemic components, similar to what we see for the vertical component of the systemic motion (see Figure~\ref{fig:LMC_velo_profile_muz0})\footnote{Animations of the variation of the morphological parameters and systemic motion in the velocity profile and maps will be made available on-line.}.  When considering literature values, and due to the strong correlation between the systemic motion and the position of the kinematic centre, we had to build the velocity maps fixing the kinematic centre to the coordinates given in the respectively works. Regarding the radial and residual tangential components, we observe strong systematic effects such as gradients across the LMC plane. The vertical velocity is systematically negative (using \citet{vandermarel02} values), or positive (using \citet{schmidt22} values) indicating that the values of centre coordinates and systemic transverse motions from the literature cannot apply to our samples. We conclude that the systematic gradients are very sensitive to small variations in the kinematic parameters. Only a narrow range of values can match the data, i.e. do not create such systematic effects, and these values are the best fit solutions given in MC21.

\begin{figure*}
    \centering
    \includegraphics[width=0.95\textwidth]{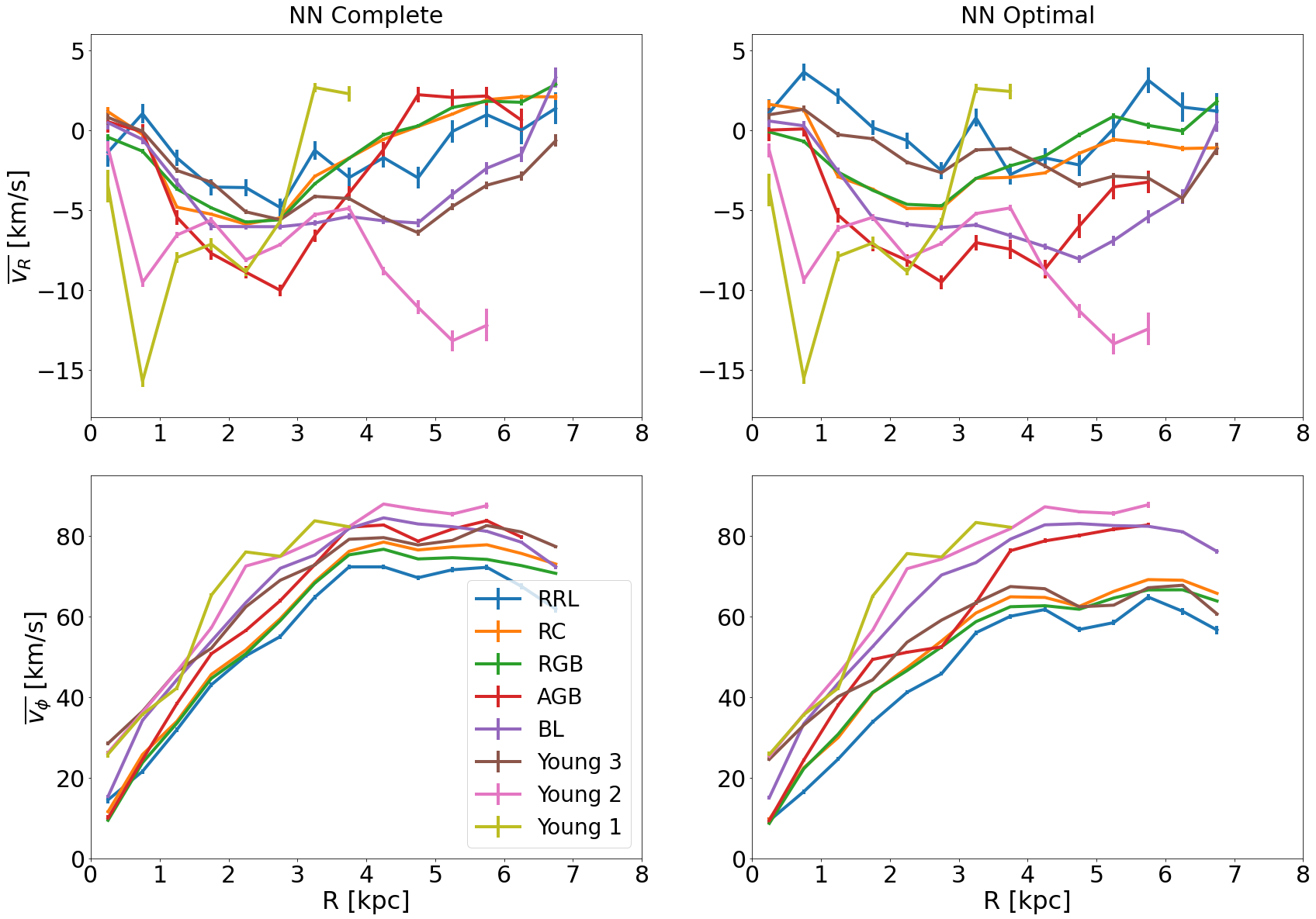}
    \caption{Stellar velocity curves of the LMC evolutionary phases. The top and bottom panels show the radial motions and rotation curves, respectively, for both NN Complete (left) and NN Optimal sample (right). Coloured lines are for the eight evolutionary phases. Only bins with more than $500$ sources are plotted.}
    \label{fig:LMC_velo_profile_pop}
\end{figure*}

Finally, in Figure~\ref{fig:LMC_velo_profile_pop}, we show the radial and tangential velocity profiles for the NN Complete (left) and NN Optimal (right) full samples separated by the same evolutionary phases selection as in MC21. We impose an additional constraint on the minimum number of $500$ sources per bin. The radial velocity profiles for the young (Young1, Young2 and Young3) samples are almost identical between the NN Complete and NN Optimal samples, so mostly not affected by MW contamination. We see how for older samples, for example for the RR Lyrae samples, the sharp minimum of velocity at $R=3\,$kpc smooths out in the outer disc and becomes more planar and even centered at zero. Differences arise in the AGB sample between the NN Complete and NN Optimal samples, oscillating as the Young1 population in the NN Complete, while remaining negative as the Young2 and Blue Loop evolutionary phases in the NN Optimal. The Young1 population highly oscillates from negative values in the inner disc to positive values at the ends of the bar. This trend might be due to a limitation in the training sample, which due to its characteristics, lacks AGB and Young1 stars. Despite this limitation, the NN classifier does an excellent job in these areas of the colour-magnitude diagram (see the two bottom rows of Fig.~\ref{fig_histogrames_classificador}, where both the AGB and Young1 areas for all LMC samples are well defined). The gradient in age observed in the tangential velocity profile in the MC21 sample is conserved in both the NN Complete and NN Optimal samples. There seems to appear a bimodality in the NN Optimal sample separating the young and old evolutionary phases, which is not present in the NN Complete sample, indicating that it might be an artifact of the imbalance between completeness and purity.  Therefore, further investigation is required in the analysis of stellar populations of the different samples.

\section{Conclusions}
\label{sec:conclusions}
In this work we analyse the velocity maps of four LMC samples, defined using different selection strategies, namely the proper motion selection, as in MC21, and three samples based on a Neural Network classification, trained using a MW+LMC simulation created by GOG. Depending on the cut on the probability of belonging to one sample or the other, $P_{cut}$, we define two samples, the more complete, NN Complete with $P_{cut}=0.01$, and the optimal, NN Optimal with $P_{cut}=0.52$, which corresponds to the optimal value based on the Receiver Operating Characteristics curve. We apply to this last sample an extra cut on the apparent $G$ magnitude of $G<19.5\,$mag, in order to remove further contamination of miss-classified faint stars. Taking advantage of the recently released spectroscopic line-of-sight velocities published in \gaianospace~DR3, we generate sub-samples that include both proper motions and line-of-sight velocities. We also adopt a new formalism in order to transform from the observable space $(\alpha,\delta,\mu_{\alpha*},\mu_{\delta},V_{los})$ to the LMC frame $(x',y',z',v_{x'},v_{y'},v_{z'})$. The advantage of this formalism based on that of \citet{vdm01} and \citet{vdm02} is the possibility to include the $V_{los}$ component when deriving internal LMC velocities.

We analyse the velocity profile and maps in the LMC coordinate system for the full samples and the $V_{los}$ sub-samples. The velocity maps corresponding to the radial and tangential component of the velocity for the PM sample are analogous to those presented in the MC21 paper, while in Section~\ref{sec:velmaps} we have analysed the differences between the samples based on a NN classification. As shown, differences are of second order and mainly located in the outer disc, where differences in density also arise. As a novelty, we include the $V_{los}$ sub-samples with line-of-sight velocities from \gaia DR3 \citep{Katz2022}, which allows the analysis of the vertical velocity component.

The main conclusions of this work are:
\begin{itemize}
\item In all samples and sub-samples, the dynamics in the inner disc are mainly bar dominated, and this is a confirmation of what was first found in MC21. An asymmetry along the bar-major axis is emphasized specially when mapping the kinematics with the $V_{los}$ sub-samples.
\item The kinematics on the spiral arm over-density seem to be dominated by an inward ($V_R<0$) motion and a rotation faster than that of the disc ($V_\phi-\bar{V_\phi}>0$) in the piece of the arm attached to the bar, though $V_{los}$ sub-samples are not conclusive in this region. 
\item The dynamics seems to change in the piece of the arm with lower density or even detached from the main arm after the density break, in the sense that the radial velocity and residual tangential velocity can reverse sign.
\item Contamination of MW stars seem to dominate the outer parts of the disc and affecting largely old and later evolutionary phase (e.g., RRL or AGB stars).
\item Uncertainties in the LMC systemic motion largely affect the vertical component of the velocity even causing a change in sign. Uncertainties in the morphological parameters of the LMC (inclination and position angle) can modify the radial and vertical velocity maps, apart from inducing a stretch or rotation of them.
\item Not having $V_{los}$ for all stars does not largely change the kinematic profiles or maps. The approximation used to derive the internal kinematics is accurate.
\end{itemize}

The available \gaianospace~DR3 dataset and the new strategy to select LMC clean samples have proven effectively suitable to perform kinematic studies and allowing a deep analysis of the nature of the LMC morphology. Comparison with realistic LMC mock catalogues are crucial and will be the goal of further work.

\section*{Acknowledgements}
 This work has made use of data from the European Space Agency (ESA) mission {\it Gaia} (\url{https://www.cosmos.esa.int/gaia}), processed by the {\it Gaia} Data Processing and Analysis Consortium (DPAC, \url{https://www.cosmos.esa.int/web/gaia/dpac/consortium}). Funding for the DPAC has been provided by national institutions, in particular the institutions participating in the {\it Gaia} Multilateral Agreement. OJA acknowledges funding by l'Agència de Gestió d'Ajuts Universitaris i de Recerca (AGAUR) official doctoral program for the development of a R+D+i project under the FI-SDUR grant (2020 FISDU 00011). OJA, MRG, XL, TA and EM acknowledge funding by the Spanish MICIN/AEI/10.13039/501100011033 and by "ERDF A way of making Europe" by the “European Union” through grant RTI2018-095076-B-C21, and the Institute of Cosmos Sciences University of Barcelona (ICCUB, Unidad de Excelencia ’Mar{\'\i}a de Maeztu’) through grant CEX2019-000918-M. TA also acknowledges the grant RYC2018-025968-I funded by MCIN/AEI/10.13039/501100011033 and by ``ESF Investing in your future''. PM acknowledges support from project grants from the Swedish Research Council (Vetenskapr\aa det, Reg: 2017-
03721; 2021-04153). LC acknowledges financial support from the Chilean Agencia Nacional de Investigaci\'{o}n y Desarrollo (ANID) through Fondo Nacional de Desarrollo Cient\'{\i}fico y Tecnol\'{o}gico (FONDECYT) Regular Project 1210992. SRF acknowledges financial support from the Spanish Ministry of Economy and Competitiveness (MINECO) under grant number AYA2016-75808-R, RTI2018-096188-B-I00, from the CAM-UCM under grant number PR65/19-22462 and the Spanish postdoctoral fellowship (2017-T2/TIC-5592).

\bibliographystyle{aa}
\bibliography{mylmcbib} 

\begin{appendix}

\section{Coordinate changes}\label{appendix_vdM}
In this appendix, we detail the steps performed to transform from heliocentric equatorial coordinates to LMC internal coordinates. This formalism is based on \citet{vdm01} and \citet{vdm02}.

\subsection{Positions}

The position of any point in space is uniquely determined by its right ascension and declination on the sky, $(\alpha, \delta)$, and its distance $D$ which can be referenced to a particular point $O$ with coordinates $(\alpha_0,\delta_0,D_0)$. We choose the LMC centre to be the reference centre.

We introduce the angular coordinates $(\phi, \rho)$ (Figure \ref{Fig1_schema}) which are defined in the celestial sphere:
\begin{itemize}
\item $\rho$ is the angular distance between the points $(\alpha, \delta)$ and $(\alpha_0,\delta_0)$.
\item $\phi$ is the position angle of the point $(\alpha,\delta)$ with respect to $(\alpha_0, \delta_0)$. In particular $\phi$ is the angle at $(\alpha_0, \delta_0)$ between the tangent to the great circle on the celestial sphere through $(\alpha, \delta)$ and $(\alpha_0, \delta_0)$, and the circle of constant declination $\delta_0$. By convention, $\phi$ is measured counterclockwise starting from the axis that runs in the direction of decreasing right ascension at constant declination $\delta_0$.
\end{itemize}

We can uniquely define $(\rho,\phi)$ as function of $(\alpha,\delta)$, for a choice of the origin $O$, by using:
\begin{equation}
\begin{split}
\rho = & \arccos \left[ \cos \delta \cos \delta_0 \cos(\alpha-\alpha_0) + \sin \delta \sin \delta_0  \right] \\
\phi = & \arctan \left[ \frac{\sin \delta \cos \delta_0 - \cos \delta \sin \delta_0 \cos (\alpha-\alpha_0)}{-\cos \delta \sin(\alpha - \alpha_0)} \right]
\end{split}
\end{equation}

These previous equations have been obtained by using the cosine, sine rule of spherical trigonometry and the so-called analogue formula:
\begin{equation}\label{eq_rhophi_alphadelta}
\begin{split}
\cos \rho & = \cos \delta \cos \delta_0 \cos (\alpha- \alpha_0) + \sin \delta \sin \delta_0 \\
\sin \rho \cos \phi & = -\cos \delta \sin (\alpha-\alpha_0) \\
\sin \rho \sin \phi & = \sin \delta \cos \delta_0 - \cos \delta \sin \delta_0 \cos (\alpha - \alpha_0)
\end{split}
\end{equation}

Then, we can introduce a Cartesian coordinate system $(x,y,z)$ that has its origin at $O$, with the x-axis anti-parallel to the right ascension axis, the y-axis parallel to the declination axis, and the z-axis towards the observer. This is somehow similar to considering the orthographic projection -a method of representing three-dimensional objects where the object is viewed along parallel lines that are perpendicular to the plane of the drawing- of the usual celestial coordinates and proper motions. This yields the following transformations:
\begin{equation}\label{eq_xyz}
\begin{split}
x & = D \sin \rho \cos \phi \\
y & = D \sin \rho \sin \phi \\
z & = D_0 - D \cos \rho
\end{split}
\end{equation}

A second Cartesian coordinate system $(x',y',z')$ is introduced. It is obtained from the system $(x,y,z)$ by counterclockwise rotation around the z-axis by an angle $\theta$, followed by a clockwise rotation around the new $x'$-axis by an angle $i$. With this definition, the $(x',y')$ plane is inclined with respect to the sky by an angle $i$ (with face-on viewing corresponding to $i=0^\circ$). The angle $\theta$ is the position angle of the line-of-nodes - the intersection of the $(x',y')$-plane and the $(x,y)$-plane of the sky-, measured counterclockwise from the $x$-axis. In practice, $i$ and $\theta$ will be chosen such that the $(x',y')$-plane coincides with the plane of the LMC disk. The transformations between the $(x',y',z')$ and the $(x,y,z)$ coordinates are:
\begin{equation}\label{eq_xyz_xyzprima}
\begin{pmatrix}
x' \\
y' \\
z'
\end{pmatrix} = 
\begin{pmatrix}
\cos \theta & \sin \theta & 0 \\
-\sin \theta \cos i & \cos \theta \cos i & -\sin i \\
-\sin \theta \sin i & \cos \theta \sin i & \cos i
\end{pmatrix}
\begin{pmatrix}
x \\
y \\
z
\end{pmatrix}
\end{equation}

One is interested in the distance $D$ of points in the $(x',y')$ plane, as function of the position $(\rho, \phi)$ on the sky. The points in this plane fulfil $z'=0$, which yields:
\begin{equation}\label{eq_dist}
D = D_{z'=0} \equiv \frac{D_0 \cos i}{\cos i \cos \rho - \sin i \sin \rho \sin (\phi-\theta)}
\end{equation}

\subsection{Velocities}\label{appendix_vel}

At any given position $(D, \rho, \phi)$ a velocity vector can be decomposed into a sum of three orthogonal components:
\begin{equation}
v_1 \equiv \frac{dD}{dt}, \hspace{1cm} v_2 \equiv D \frac{d \rho}{dt}, \hspace{1cm} v_3 \equiv D \sin \rho \frac{d \phi}{dt}
\end{equation}

Here, $v_1$ is the line-of-sight velocity and $v_2$ and $v_3$ are the velocity components in the plane of the sky. Computing the time derivative of Eq. (\ref{eq_xyz}) yields:
\begin{equation}\label{eq_vxvyvz_v1v2v3}
\begin{pmatrix}
v_x \\
v_y \\
v_z
\end{pmatrix} = 
\begin{pmatrix}
\sin \rho \cos \phi & \cos \rho \cos \phi & -\sin \phi \\
\sin \rho \sin \phi & \cos \rho \sin \phi & \cos \phi \\
-\cos \rho & \sin \rho & 0
\end{pmatrix}
\begin{pmatrix}
v_1 \\
v_2 \\
v_3
\end{pmatrix}
\end{equation}

where $(v_x, v_y, v_z)$ is the three-dimensional velocity in the $(x,y,z)$ coordinate system. Again, to describe the internal kinematic of the galaxy it is useful to adopt the second Cartesian coordinate system $(x',y',z')$. We remind that the $(x',y')$-plane coincides with the plane of the LMC disk. Upon taking the time derivative on both sides on Eq. (\ref{eq_xyz_xyzprima}) yields the transformation equations from $(v_x,v_y,v_z)$ to $(v_x',v_y',v_z')$. This result can be used with Eq. (\ref{eq_vxvyvz_v1v2v3}) to obtain:
\begin{equation}
\begin{split}
\begin{pmatrix}
v_x' \\
v_y' \\
v_z'
\end{pmatrix} = &
\begin{pmatrix}
\cos \theta & \sin \theta & 0  \\
-\sin \theta \cos i & \cos \theta \cos i & -\sin i \\
-\sin \theta \sin i & \cos \theta \sin i & \cos i
\end{pmatrix}
\times \\
& \times
\begin{pmatrix}
\sin \rho \cos \phi & \cos \rho \cos \phi & -\sin \phi \\
\sin \rho \sin \phi & \cos \rho \sin \phi & \cos \phi \\
-\cos \rho & \sin \rho & 0
\end{pmatrix}
\begin{pmatrix}
v_1 \\
v_2 \\
v_3
\end{pmatrix}
\end{split}
\label{eq_vlos}
\end{equation}

We know that $v_1$ is the line-of-sight velocity. Now, we need to relate the velocities $v_2$ and $v_3$ with the proper motions $\mu_{\alpha*}$ and $\mu_\delta$. In these directions the proper motions are defined as:
\begin{equation}
\mu_{\alpha*} \equiv \cos \delta \frac{d \alpha}{d t} \hspace{0.1cm} , \hspace{0.6cm}  \mu_\delta \equiv \frac{d \delta}{d t}
\end{equation}

Upon taking time derivative of Eq. (\ref{eq_rhophi_alphadelta}) we obtain relations between $d\rho/dt$ and $d\phi/dt$ on the one hand and $d\alpha/dt$ and $d\delta/dt$ on the other hand. This system can be solved to obtain:
\begin{equation}
\begin{pmatrix}
v_2 \\
v_3 
\end{pmatrix} =
D
\begin{pmatrix}
\sin \Gamma & \cos \Gamma \\
\cos \Gamma & - \sin \Gamma 
\end{pmatrix}
\begin{pmatrix}
\mu_{\alpha*} \\
\mu_\delta
\end{pmatrix}
\end{equation}

where the angle $\Gamma$ determines the rotation angle of the $(v_2, v_3)$  frame on the sky. It is given by:
\begin{equation}
\begin{split}
\cos \Gamma & = [\sin \delta \cos \delta_0 \cos (\alpha -\alpha_0) - \cos \delta \sin \delta_0] / \sin \rho \\
\sin \Gamma & = [\cos \delta_0 \sin (\alpha -\alpha_0)] / \sin \rho 
\end{split}
\end{equation}

\subsubsection{Correcting from the systemic motion}\label{append_sys_motion}

For a planar system, the velocity of a tracer can be written as a sum of three components: the velocity corresponding to the systemic motion, the velocity corresponding to precession and nutation of the disk plane and the velocity corresponding to the internal motion of the tracer: 
\begin{equation}
\begin{pmatrix}
v_1 \\
v_2 \\
v_3
\end{pmatrix} =
\begin{pmatrix}
v_1 \\
v_2 \\
v_3
\end{pmatrix}_{sys} +
\begin{pmatrix}
v_1 \\
v_2 \\
v_3
\end{pmatrix}_{pn} + 
\begin{pmatrix}
v_1 \\
v_2 \\
v_3
\end{pmatrix}_{int}
\end{equation}

Then, if we neglect the effect of precession and nutation, we can determine the internal motion by using:
\begin{equation}\label{eq:v123_int}
\begin{pmatrix}
v_1 \\
v_2 \\
v_3
\end{pmatrix}_{int} =
\begin{pmatrix}
v_1 \\
v_2 \\
v_3
\end{pmatrix} -
\begin{pmatrix}
v_1 \\
v_2 \\
v_3
\end{pmatrix}_{sys}
\end{equation}

where we have explained how to compute the first term in Section \ref{appendix_vel}. On the other hand, we can determine the systemic motion using the inverse relation of Eq. (\ref{eq_vxvyvz_v1v2v3}):

\begin{equation}
\begin{split}
\begin{pmatrix}
v_1 \\
v_2 \\
v_3
\end{pmatrix}_{sys} = &
\begin{pmatrix}
\sin \rho \cos \phi & \sin \rho \sin \phi & -\cos \rho \\
\cos \rho \cos \phi & \cos \rho \sin \phi & \sin \rho \\
-\sin \phi & \cos \phi & 0
\end{pmatrix}
\begin{pmatrix}
v_x \\
v_y \\
v_z
\end{pmatrix}_{sys} = \\
= & D_0 \begin{pmatrix}
\sin \rho \cos \phi & \sin \rho \sin \phi & -\cos \rho \\
\cos \rho \cos \phi & \cos \rho \sin \phi & \sin \rho \\
-\sin \phi & \cos \phi & 0
\end{pmatrix}
\begin{pmatrix}
\mu_{x,0} \\
\mu_{y,0} \\
\mu_{z,0}
\end{pmatrix}
\end{split}
\end{equation}

where $\mu_{x,0}$ and $\mu_{y,0}$ are the associated proper motions in the $x$ and $y$ directions at the centre of the disc, and $\mu_{z,0}=v_{z,0}/D_{0}$, the associated line-of-sight velocity, expressed on the same scale as the proper motions by dividing by $D_{0}$.

\subsection{Estimating the observational line-of-sight velocity}\label{append_no_vlos}

In case we do not have observational information on the line-of-sight velocity we can estimate $v_{1,int}$ by computing the derivative of the distance as function of time:
\begin{equation}\label{eq:v1_int}
\begin{split}
v_{1,int} \equiv & \frac{dD}{dt} = \frac{d}{dt} \left[ \frac{D_0 \cos i}{\cos i \cos \rho - \sin i \sin \rho \sin (\phi-\theta)} \right] \\
= & \frac{v_{2,int} [\cos i \sin \rho + \sin i \cos \rho \sin (\phi - \theta)] + v_{3,int} \sin i \cos (\phi - \theta)}{\cos i \cos \rho - \sin i \sin \rho \sin (\phi - \theta)}
\end{split}
\end{equation}

To compute $v_{2,int}$ and $v_{3,int}$ when not having observational information on the line-of-sight velocity we proceed the same way it is detailed in Section \ref{appendix_vel} since $v_{2,int}$ and $v_{3,int}$ only depend on the proper motions.

\end{appendix}
\label{lastpage}

\end{document}